\documentclass{aa}

\usepackage{microtype}
\usepackage{hyperref}
\usepackage{booktabs}
\usepackage{multirow, bigdelim}
\usepackage{bm}
\usepackage{subcaption}
\usepackage{tablefootnote}
\usepackage{threeparttable}
\usepackage{graphicx}
\usepackage{siunitx}

\hypersetup{%
  colorlinks=true,
  linkcolor=blue,
  citecolor=blue
}

\begin{document}

\title{Evolutionary view through the starless cores in Taurus}

\subtitle{Deuteration in TMC~1-C and TMC~1-CP}

\author{D. Navarro-Almaida\inst{1}
	\and
	A.~Fuente\inst{1}
	\and
	L. Majumdar\inst{2}
    \and
    V. Wakelam\inst{3}
    \and
    P. Caselli\inst{4}
 	\and
    P. Rivi\`ere-Marichalar\inst{1}
    \and
    S. P. Trevi\~no-Morales\inst{5}
    \and
    S. Cazaux \inst{6}
    \and   
    I. Jim\'enez-Serra\inst{7}
    \and
    C. Kramer\inst{8}
    \and 
    A. Chac\'on-Tanarro\inst{1}
    \and
    J. M. Kirk\inst{9}
    \and
    D.~Ward-Thompson\inst{9}
    \and
    M.~Tafalla\inst{1}
    }
  
\institute{ 
	Observatorio Astron\'omico Nacional (OAN), Alfonso XII, 3,  28014, Madrid, Spain
    \and
    School of Earth and Planetary Sciences, National Institute of Science Education and Research, HBNI, Jatni 752050, Odisha, India
	\and
	Laboratoire d'Astrophysique de Bordeaux, Univ. Bordeaux, CNRS, B18N, all\'ee Geoffroy Saint-Hilaire, 33615 Pessac, France
	\and
	Centre for Astrochemical Studies, Max-Planck-Institute for Extraterrestrial Physics, Giessenbachstrasse 1, 85748, Garching, Germany
	\and
	Chalmers University of Technology, Department of Space, Earth and Environment, SE-412 93 Gothenburg, Sweden
	\and
	Faculty of Aerospace Engineering, Delft University of Technology, Delft, The Netherlands; University of Leiden, P.O. Box 9513, NL, 2300 RA, Leiden, The Netherlands
	\and
	Centro de Astrobiolog\'{\i}a (CSIC-INTA), Ctra. de Ajalvir, km 4, Torrej\'on de Ardoz, 28850, Madrid, Spain
	\and
	Institut de Radio Astronomie Millim\'etrique (IRAM), 300 Rue de la Piscine, Domaine Universitaire, 38406 Saint Martin d'H\`{e}res, France
	\and
	Jeremiah Horrocks Institute, University of Central Lancashire, Preston PR1 2HE, UK
}

\date{}

\abstract {The chemical and physical evolution of starless and pre-stellar cores are of paramount importance to understanding the process of star formation. The Taurus Molecular Cloud cores TMC~1-C and TMC~1-CP share similar initial conditions and provide an excellent opportunity to understand the evolution of the pre-stellar core phase.}
{We investigated the evolutionary stage of starless cores based on observations towards the prototypical dark cores TMC~1-C and TMC~1-CP.}
{We mapped the prototypical dark cores TMC~1-C and TMC~1-CP in the CS $3\rightarrow 2$, C$^{34}$S $3\rightarrow 2$, $^{13}$CS $2\rightarrow 1$, DCN $1\rightarrow 0$, DCN $2\rightarrow 1$, DNC $1\rightarrow 0$, DNC $2\rightarrow 1$,  DN$^{13}$C $1\rightarrow 0$, DN$^{13}$C $2\rightarrow 1$, N$_2$H$^+$ $1\rightarrow 0$, and N$_2$D$^+$ $1\rightarrow 0$ transitions. We performed a multi-transitional study of CS and its isotopologs, DCN, and DNC lines to characterize the physical and chemical properties of these cores. We studied their chemistry using the state-of-the-art gas-grain chemical code \textsc{Nautilus} and pseudo time-dependent models to determine their evolutionary stage.}
{The central $n_{\rm H}$ volume density, the N$_{2}$H$^{+}$ column density, and the abundances of deuterated species are higher in TMC~1-C than in TMC~1-CP, yielding a higher N$_2$H$^+$ deuterium fraction in TMC~1-C, thus indicating a later evolutionary stage for TMC~1-C. The chemical modeling with pseudo time-dependent models and their radiative transfer are in agreement with this statement, allowing us to estimate a collapse timescale of $\sim 1$ Myr for TMC~1-C. Models with a younger collapse scenario or a collapse slowed down by a magnetic support are found to more closely reproduce the observations towards TMC~1-CP.}
{Observational diagnostics seem to indicate that TMC~1-C is in a later evolutionary stage than TMC~1-CP, with a chemical age $\sim$1 Myr. TMC~1-C shows signs of being an evolved core at the onset of star formation, while TMC~1-CP appears to be in an earlier evolutionary stage due to a more recent formation or, alternatively, a collapse slowed down by a magnetic support.}

\keywords{Astrochemistry --
  ISM: kinematics and dynamics --
  ISM: abundances --
  ISM: molecules -- Stars: formation -- Stars: low-mass}
\maketitle

\section{Introduction}
\label{sec:intro}

Starless cores are the seeds of star formation, and are the result of fragmentation and contraction of molecular clouds. A detailed knowledge of the physical and chemical conditions in a starless core is then crucial to understand the subsequent stellar evolution. Starless cores are characterized by densities of around $n_{\rm H} = 10^{5}$ cm$^{-3}$, and cold ($T \leq 10$ K) temperatures \citep{WardThompson1996, Tafalla2004, Crapsi2005, Keto2008}. In these particular physical conditions, species like CO, are expected to be heavily depleted from the gas, being frozen out onto dust grain surfaces \citep{Leger1983}, and thus becoming inadequate tracers of mass, hydrogen number density $n_{\rm H}$, and kinetic temperature $T_{\rm kin}$. Fortunately, some species remain abundant even in the dense and cold inner regions of starless cores. One of these cases is molecular ion N$_2$H$^+$, rapidly destroyed through reactions with CO in molecular clouds. Once CO is depleted, the N$_2$H$^+$ abundance increases, remaining in the gas phase even at high densities. The anti-correlation between N$_{2}$H$^{+}$ and CO line emissions \citep[see, e.g.,][]{Pagani2007, Caselli1999}, and the high critical density of N$_{2}$H$^{+}$ $1\rightarrow0$ line \citep[$n_{\rm crit}\sim 1.5\times 10^{5}$ cm$^{-3}$ at 10 K, ][]{Lin2020}, make this molecule suitable to probe inside dense cores.

Deuterated compounds are also thought to be good tracers of the cold gas inside dense starless cores \citep{Millar1989}. At the low temperatures prevailing in these environments, the exothermicity of the reaction
\begin{equation}\label{eq:deut}
	{{\rm H}_{3}}^{+} + {\rm HD} \leftrightharpoons {\rm H}_{2}{\rm D}^{+} + {\rm H}_{2}
\end{equation}
in the forward direction \citep[of $\sim 232$ K, see][]{Gerlich2002} promotes the formation of ${\rm H}_{2}{\rm D}^{+}$ and inhibits the reverse reaction. The deuteration proceeds further via formation of D$_2$H$^+$ and D$_3^+$ in reactions with HD and D$_2$. These deuterated ions react with a variety of neutral molecules, such as CO and N$_2$, leading to high abundances of other deuterated species \citep[see, e.g.,][]{Ceccarelli2014, Roueff2014}. As a result, the deuteration fraction, which is the abundance of a given deuterated compound relative to its hydrogenated counterpart, can be enhanced by more than three orders of magnitude \citep[see, e.g.,][]{Turner2001, Bacmann2003, Crapsi2005, Pagani2007, Spezzano2013} compared to the solar elemental D/H ratio \citep[$1.51^{+0.39}_{-0.33}\times 10^{-5}$, see, e.g.,][]{Linsky2003, Oliveira2003}. Deuterated molecules thus become important diagnostic tools of dense and cold interstellar clouds.

The N$_{2}$H$^{+}$ deuteration fraction (the ratio of the column density of N$_{2}$D$^{+}$ to its non-deuterated counterpart N$_{2}$H$^{+}$) has been found in moderate anti-correlation with the gas temperature \citep{Emprechtinger2009, Chen2011, Fontani2015} and in correlation with the density \citep{Daniel2007}. These dependencies are also tightly related to the evolutionary stage of the object. Using a chemical model, \citet{Caselli2002} concluded that the N$_{2}$D$^{+}$/N$_{2}$H$^{+}$ column density ratio is a chemical clock, working as an indicator for the evolutionary stage in low-mass star formation sites. Increasing attention is being paid to the accurate modeling of the deuterium chemistry as a tool to probe the evolution of the dense gas during the starless core phase \citep[see, e.g.,][]{Sipila2010, Sipila2015, Roueff2015, Majumdar2017}. One key aspect is the realization of the importance of the spin state in deuterium chemistry \citep{Sipila2019}, since multiply-hydrogenated or deuterated molecules can exist in several forms due to the different nuclear spin states. \citet{Majumdar2017} presented the first publicly available chemical network for \textsc{Nautilus} \citep{Ruaud2016} with deuterated species and spin chemistry in its two-phase implementation where the gas phase and grain phase interact. 

\begin{figure*}
	\centering
		\includegraphics[width=0.99\textwidth,keepaspectratio]{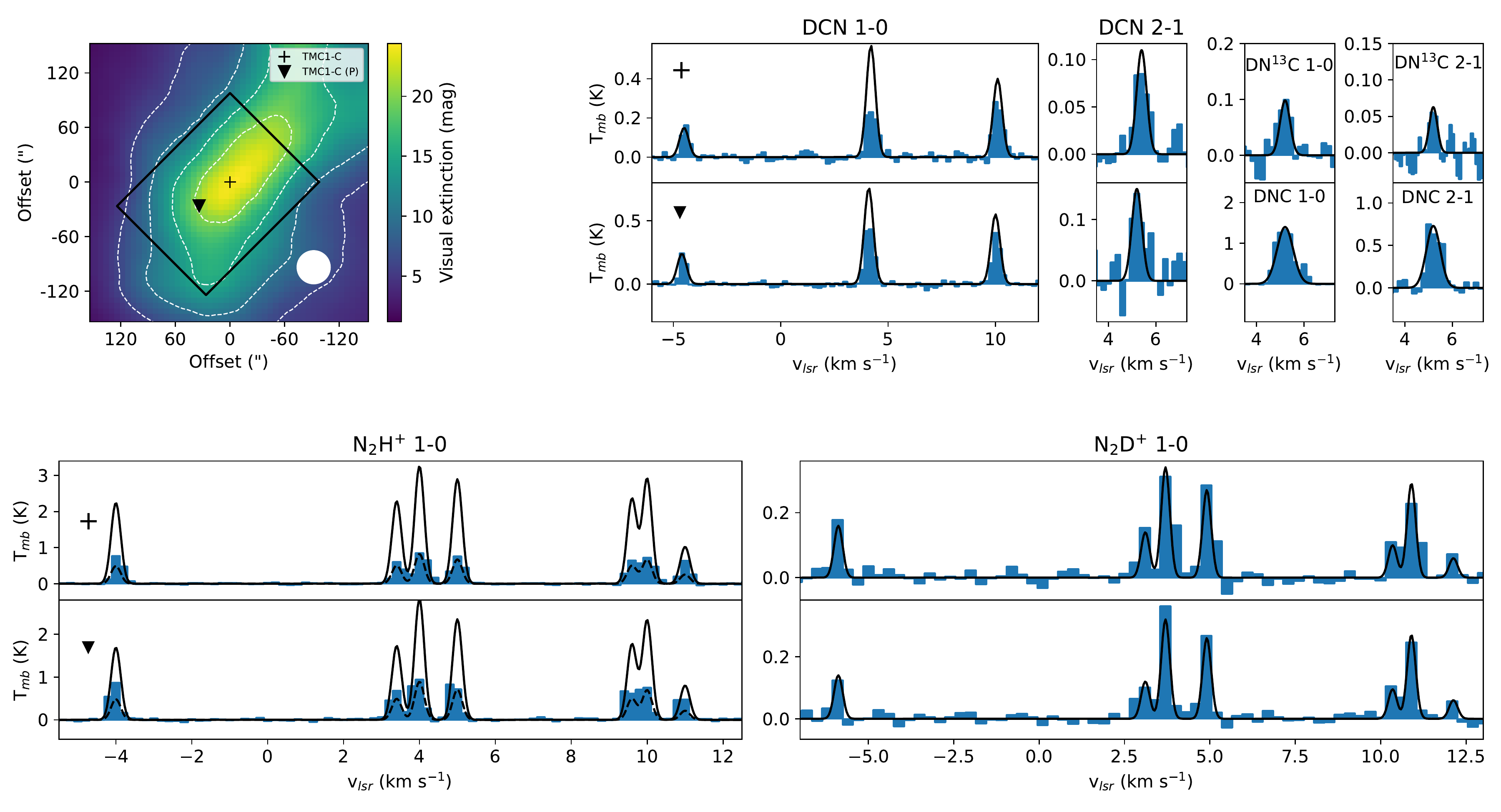}
		$\left.\right.$\\$\left.\right.$
		\includegraphics[width=\textwidth,keepaspectratio]{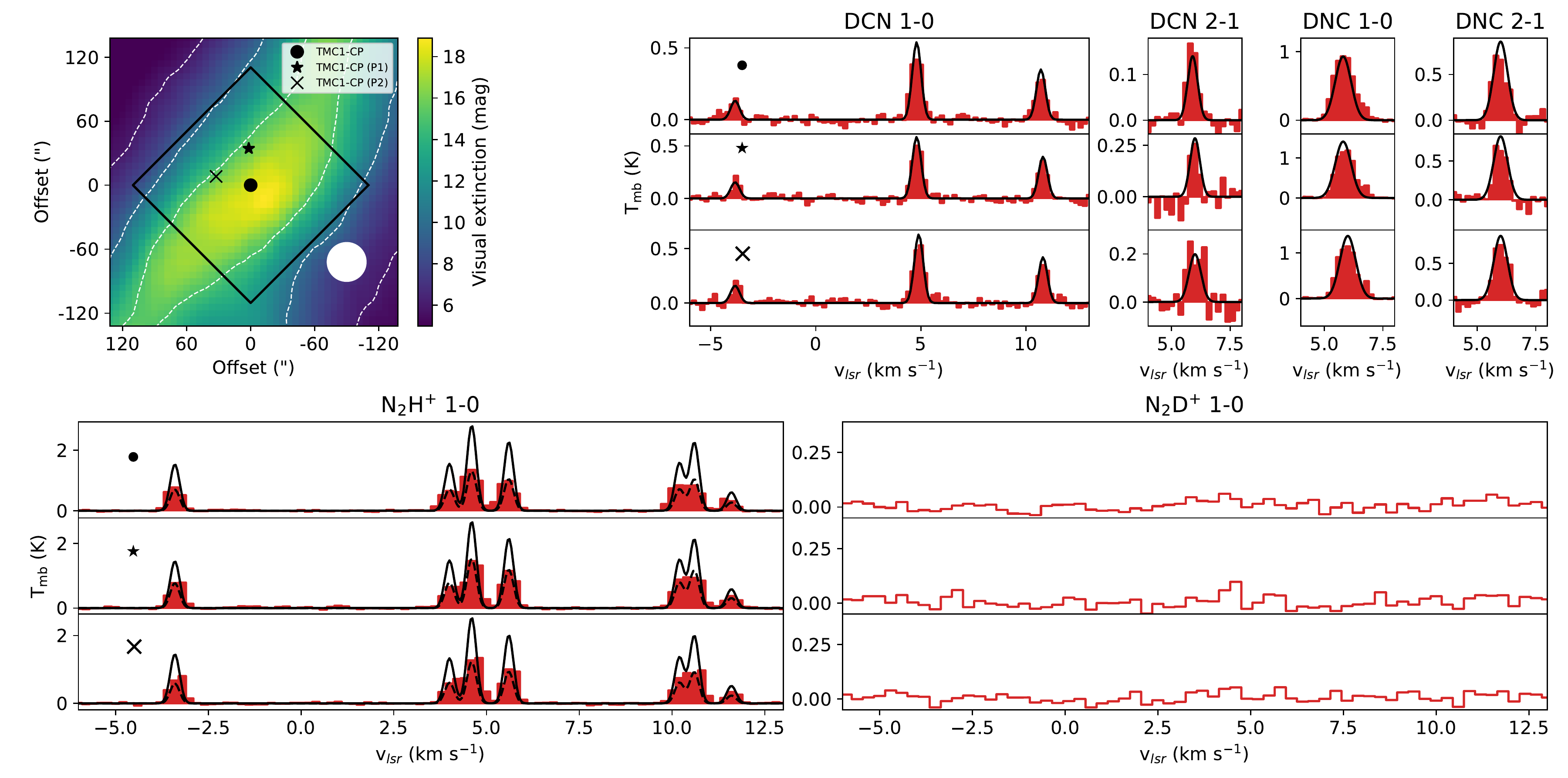}
		\caption{\emph{Top left and third row, left:} Visual extinction maps from Herschel data (Jason Kirk, private communication) of the TMC~1-C and TMC~1-CP cores, respectively, with the areas mapped in the transitions presented in Table \ref{tab:summarylines}. Contour lines show levels of extinction of 6, 10, 15, and 20 mag. The blue and red spectra show the observed molecular lines in TMC~1-C and TMC~1-CP, respectively. The solid lines show synthetic spectra created with the model parameters present in Table~\ref{tab:col_dens}.}
		\label{fig:spectra}
\end{figure*}

The chemical properties of star-forming regions are best constrained with chemical models of well-known objects. The Taurus molecular cloud (TMC), at a distance of $141.8\pm 0.2$ pc \citep{Galli2018}, is one of the closest molecular cloud complexes and is considered an archetypal low-mass star-forming region. The most massive molecular cloud in Taurus is the Heiles cloud 2 (HCL 2), which hosts the well-known region TMC~1. TMC~1-C and TMC~1-CP are two starless dense cores embedded in the translucent filament TMC~1 (see Figure~\ref{fig:spectra}). These two nearby cores have similar masses ($\sim$ 1~M$_\odot$), gas kinetic temperatures of 8$-$10 K (Kirk et al., in prep.), and share similar initial conditions. Moreover, \citet{NavarroAlmaida2020} obtained a similar density profile for the two cores from a multi-transition analysis of CS and its isotopologs. In spite of these similarities, several spectroscopic studies points to significant chemical and dynamical differences between them. TMC~1-CP has been the target of numerous chemical studies and is considered as prototype of a C-rich (C/O$\sim$1) starless core \citep{Feher2016, Gratier2016, Agundez2013}. With a rich carbon chemistry, it is the preferred site for hunting new organic molecules \citep{Guire2020, Lee2021}. Less studied from the chemical point of view, TMC~1-C has been identified as an accreting starless core with high depletion of CO \citep{Schnee2007, Schnee2010}. In addition to CO, other dense gas tracers like CS are expected to be depleted in dense cores \citep{Kim2020}. In line with this statement, \citet{Fuente2019} measured a higher depletion of CS towards TMC~1-C, based on data from the Gas phase Elemental abundances in Molecular CloudS (GEMS) IRAM 30m Large Program. One compelling explanation of these measurements is that the differences are due to different evolutionary stages of the two cores \citep[e.g.,][]{Sipila2018}. If confirmed, these nearby cores would constitute a valuable site to investigate the evolution of gas and dust during the pre-stellar phase, avoiding the confusion due to possible differences in the initial conditions or environment when using cores placed in different molecular clouds. 

In this paper we investigate the evolutionary stage of TMC~1-C and TMC~1-CP based on spectroscopic observations. Several observational diagnostics have been proposed to determine the evolutionary stage of starless and pre-stellar cores \citep{Crapsi2005}, among which the volume density at the core center, the N$_2$H$^+$ column density at the visual extinction peak, the depletion of CO and CS, and the deuterium fraction of N$_2$H$^+$. We mapped the two cores in several lines of CS, N$_2$H$^+$, N$_2$D$^+$, DCN, and DNC to derive an accurate estimate of the central density. Moreover, we used the CS/N$_2$H$^+$ and N$_{2}$D$^{+}$/N$_{2}$H$^{+}$, and DCN/DNC column density ratios as chemical clocks for the core evolution. Finally, we performed pseudo-dynamical modeling and 3D radiative transfer to synthesize the spectra towards TMC~1-C to compare with observations.

\section{Observational strategy}
\label{sec:obs}

\begin{table}
	\centering
	\caption{Observed lines, beam sizes, and efficiencies}\label{tab:summarylines}
	\resizebox{0.499\textwidth}{!}{
	\begin{tabular}{rrcc}
		\toprule
		Line &  Frequency (MHz)$^{a}$ & $\theta_{\rm beam}$(")$^{b}$ & F$_{\rm eff}$/B$_{\rm eff}$ \\ \midrule
				$^{13}$CS $2\rightarrow1$ & 92494.28 & 27 & 0.95/0.81 \\
				C$^{34}$S $3\rightarrow2$ & 144617.10 & 17 & 0.93/0.74 \\
				CS $3\rightarrow2$ & 146969.03 & 17 & 0.93/0.73 \\
				DN$^{13}$C $1\rightarrow0$ & 73367.75 & 34 & 0.95/0.83\\
				DN$^{13}$C $2\rightarrow1$ & 146734.00 & 17 & 0.93/0.73\\
				DNC $1\rightarrow0$ & 76305.68 & 32 & 0.95/0.83\\
				DNC $2\rightarrow1$ & 152609.74 & 16 & 0.93/0.72\\
				DCN $1\rightarrow0$ & 72414.93 & 34 & 0.95/0.83\\
				DCN $2\rightarrow1$ & 144828.11 & 17 & 0.93/0.74\\
				N$_{2}$H$^{+}$ $1\rightarrow0$ & 93173.38 & 26 & 0.95/0.81\\
				N$_{2}$D$^{+}$ $1\rightarrow0$ & 77109.61 & 32 & 0.95/0.83\\
    		\bottomrule
	\end{tabular}
	}
	\flushleft
			{\small
			\ \ \ \ $^{\bm{(a)}}$ When hyperfine splitting is present, the frequency shown here is that of the main component. The frequencies are taken from the CDMS catalog except those of the DN$^{13}$C lines, taken from SLAIM catalog.\\
			\ \ \ \ $^{\bm{(b)}}$ The HPBW(") is computed as 2460/Freq(GHz), see \href{https://publicwiki.iram.es/Iram30mEfficiencies}{IRAM 30m Efficiencies}\\
			}
\end{table}

\begin{table}
	\centering
	\caption{Target positions}\label{tab:targets}
	\resizebox{0.499\textwidth}{!}{
	\begin{tabular}{lccc}
		\toprule
		Source &  RA (2000) &  Dec (2000) & $v_{\rm{LSR}}$ (km s$^{-1}$) \\ \midrule
				TMC~1-C & 04:41:37.58 & 26:00:31.10 & 5.2\\
				TMC~1-C (P) & 04:41:40.10 & 26:00:05.00 & 5.2\\
				TMC~1-CP & 04:41:41.90 & 25:41:27.10 & 5.8 \\
				TMC~1-CP (P1) & 04:41:42.56 & 25:42:02.40 & 5.8\\
				TMC~1-CP (P2) & 04:41:44.07 & 25:41:42.43 & 5.8\\
    		\bottomrule
	\end{tabular}
	}
\end{table}

In order to determine the morphology of the dense gas and the peak hydrogen densities in TMC~1-C and TMC~1-CP, we mapped these starless cores in a selection of molecular lines (see Table~\ref{tab:summarylines}) using the IRAM 30m telescope (Pico Veleta, Spain). The observations were carried out during May 2019 using the on-the-fly (OTF) technique to cover a field of view of 9 and 6.8 arcmin$^{2}$ towards TMC 1-C and TMC 1-CP, respectively. We used 2 mm and 3 mm bands in dual polarization mode using the EMIR receivers \citep{Carter2012} with the Fast Fourier Transform Spectrometer (FFTS) at 50 kHz resolution. The emission from the sky was subtracted using reference positions located at offsets (600'', 0'') relative to the map centers ($\alpha$ = 04$^h$41$^m$37$^s$58, $\delta$= 26$^\circ$00'31''.10 for  TMC~1-C; $\alpha$ = 04$^h$41$^m$40$^s$.10, $\delta$ = 26$^\circ$00$'$05$''$.00 for TMC~1-CP, see Table \ref{tab:targets}). In addition to the maps, long integration spectra were taken towards the map centers to improve the S/N of the 2~mm lines. In this paper we use the main beam brightness temperature $T_{\rm MB}$ as intensity scale, while the output of the telescope is calibrated in antenna temperature $T_{\rm A}$. The conversion between $T_{\rm MB}$ and $T_{\rm A}$ is done using the Ruze equation utility provided in the data reduction and line analysis package GILDAS-CLASS\footnote{GILDAS home page: \href{https://www.iram.fr/IRAMFR/GILDAS/}{www.iram.fr/IRAMFR/GILDAS/}}.

\section{Results}\label{sec:results}

\begin{figure*}
	\centering
		\includegraphics[width=0.95\textwidth,keepaspectratio]{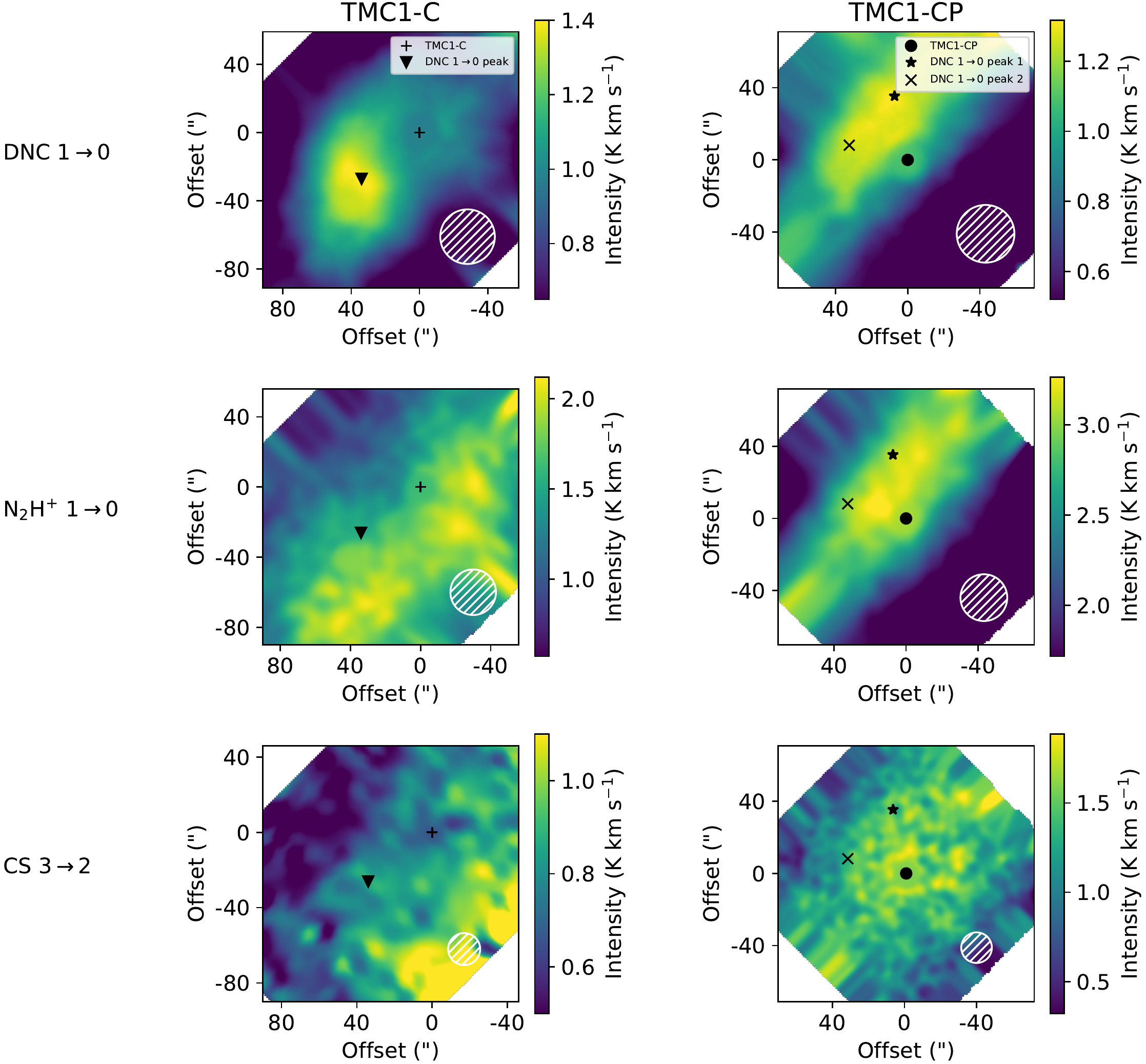}
		\caption{Velocity-integrated intensity maps of TMC~1-C (left) and TMC~1-CP (right) molecular emission of DNC $1\rightarrow 0$, N$_{2}$H$^{+}$ $1\rightarrow 0$, and CS $3\rightarrow 2$. The integration was performed from 4 to 8 km s$^{-1}$ when no hyperfine structure is present. Otherwise, the emission is integrated to cover all hyperfine components. Also shown are the selected positions from Table \ref{tab:targets} and the beam sizes from Table \ref{tab:summarylines} in each map.}
		\label{fig:maps}
\end{figure*}

Figure~\ref{fig:spectra} shows the visual extinction maps towards the starless cores TMC~1-C and TMC~1-CP as derived from Herschel maps (J. Kirk, private communication). The overplotted black rectangles indicate the areas mapped in molecular lines with the 30m telescope. The spectra from the positions in Table \ref{tab:targets} are also shown. The map center positions of TMC~1-C and TMC~1-CP correspond to the Herschel extinction peaks of each core (see extinction maps in Figure~\ref{fig:spectra}). We also included the spectra towards the DNC $1\rightarrow0$ emission peak positions TMC~1-C (P), TMC~1-CP (P1), and TMC~1-CP (P2) to compare with the map centers (see Figure~\ref{fig:maps}). It should be noted that the N$_2$D$^+$ $1\rightarrow0$, DN$^{13}$C $1\rightarrow0$, and DN$^{13}$C $2\rightarrow1$ lines were only detected towards TMC~1-C. The integrated areas and main-beam temperatures of the observed lines are shown in Table~\ref{tab:lines1} to \ref{tab:lines5}.

The integrated intensity maps of the DNC $1\rightarrow 0$, N$_{2}$H$^{+}$ $1\rightarrow 0$, and the CS $3\rightarrow 2$ lines towards the studied cores are shown in Figure~\ref{fig:maps}. Integration was performed over the whole hyperfine structure for those lines that present it. The DNC $2\rightarrow1$, DCN $2\rightarrow1$, N$_2$D$^+$ $1\rightarrow0$, DN$^{13}$C $1\rightarrow0$, and DN$^{13}$C $2\rightarrow 1$ lines were too weak to produce maps. In Figure~\ref{fig:maps}, we present maps of integrated intensity of different lines to analyze their morphology. Although the discussion of emission morphology from optically thick lines is, in general, uncertain, the morphologies of the integrated intensity maps shown in Figure~\ref{fig:maps}, which include integration over all hyperfine components, are similar to those that only include integration over the weakest optically thin hyperfine component, and thus the following analysis is not affected by opacity. The emission of the CS $3\rightarrow2$ line is quite uniform near the position TMC~1-C. This flat map is expected because the CS $3\rightarrow2$ line is optically thick in this dense core. Moreover, \citet{Fuente2019} found that the CS abundance is depleted towards the visual extinction peak. The N$_2$H$^{+}$ $1\rightarrow0$ emission presents an elongated structure following the direction of the molecular filament. The bright area in the N$_2$H$^+$ $1\rightarrow0$ emission is not coincident with the location of the the visual extinction peak, but shifted $\sim$10$"$ SW (see visual extinction maps in Figure~\ref{fig:spectra}). Contrary to CS and N$_2$H$^+$, the morphology of the DNC $1\rightarrow0$ map resembles that observed in the visual extinction maps (Figure~\ref{fig:spectra}). In spite of this, the position of the emission peak of the DNC $1\rightarrow0$ line in TMC~1-C, called TMC~1-C (P), is not located towards the position of the extinction peak, as measured with the Herschel extinction maps at an angular resolution of $\sim 30$", but to the south. Interestingly, this position is $\sim 19$" SW  away (within the DNC $1\rightarrow0$ line beam size) from the visual extinction peak derived by \citet{Schnee2005} based on SCUBA observations of the dust continuum emission at 450 $\mu$m (half power beam width HPBW $\sim$7.5$"$) and 850 $\mu$m (HPBW$\sim$14$"$), and therefore the effective core center.

In TMC~1-CP, the emission of the CS $3\rightarrow2$ presents a featureless structure, wide and flat on the top, consistent with the emission of this line being optically thick. Similarly to DNC, the N$_2$H$^+$ emission is more intense in an elongated structure parallel to the dust filament. The DNC filament is farther from the visual extinction filament than N$_2$H$^+$, forming some kind of DNC--N$_2$H$^+$ layered structure. In addition to the map center, we selected two positions, TMC~1-CP (P1) and TMC~1-CP (P2), for their detailed study.  

In the two cores the morphologies observed in the DNC $1\rightarrow0$ line are more similar to those observed in the visual extinction maps. In TMC~1-C, the emission of DNC shows a rounded core, with an intense peak in the southwest part of the core. In TMC~1-CP, the DNC $1\rightarrow 0$ emission is elongated, following the filament similarly to N$_2$H$^+$, but shifted as a function of radius with respect to the main axis of the core. Assuming that DNC is coming from the denser gas, the DNC $1\rightarrow0$ images point to a difference in the dense gas morphology between the TMC~1-C and TMC~1-CP cores. The  more compact distribution of the dense gas in TMC~1-C argues in favor of this core being in a more evolved stage.

\section{Molecular column densities}

\begin{table*}
	\centering
	\caption{Kinetic temperatures, hydrogen nuclei number densities, and column densities of the species resulting from the multi-transition analysis described in Section 4.}
	\resizebox{\textwidth}{!}{
	\begin{tabular}{llccccc}
		\toprule
		&  & TMC~1-C & TMC~1-C (P) & TMC~1-CP & TMC~1-CP (P1) & TMC~1-CP (P2) \\ \midrule
			&T$_{k}$(K) $^{\bm{(a)}}$ & $\bm{7.4\pm 2.1}$ & $\bm{8.7\pm 1.9}$ & $\bm{9.0\pm 2.1}$ & $\bm{9.3\pm 1.8}$ & $\bm{9.1\pm 1.6}$\\
			CS & n$_{\rm H}$ (cm$^{-3}$) & $(1.4\pm 0.8)\times 10^{5}$ & $(8.8\pm 6.0)\times 10^{4}$ & $(6.2\pm 2.0)\times 10^{4}$ & $(6.7\pm 2.5)\times 10^{4}$ & $(7.0\pm 4.6)\times 10^{4}$ \\
			& N (cm$^{-2}$)& $(1.0\pm 0.5)\times 10^{14}$ &	 $(1.2\pm 0.9)\times 10^{14}$ & $(2.9\pm 1.8)\times 10^{14}$ & $(1.6\pm 1.2)\times 10^{14}$ & $(2.5\pm 0.9)\times 10^{14}$\\ \midrule
			\multirow{2}{*}{DNC} & n$_{\rm H}$ (cm$^{-3}$) & $(1.8\pm 0.5)\times 10^{5}$ & $\bm{(2.4\pm 0.5)\times 10^{5}}$ & $\bm{(4.0\pm 1.0)\times 10^{5}}$ & $\bm{(3.0\pm 0.6)\times 10^{5}}$ & $\bm{(4.0\pm 0.8)\times 10^{5}}$ \\
			& N (cm$^{-2}$)& $(3.0\pm 0.9)\times 10^{12}$ &	 $(3.2\pm 0.7)\times 10^{12}$ & $(2.0\pm 0.5)\times 10^{12}$ & $(2.8\pm 0.5)\times 10^{12}$ & $(2.5\pm 0.4)\times 10^{12}$\\ \midrule
			\multirow{2}{*}{DN$^{13}$C} & n$_{\rm H}$ (cm$^{-3}$) & $\bm{(4.1\pm 1.2)\times 10^{5}}$ & \multirow{2}{*}{$-$} & \multirow{2}{*}{$-$} & \multirow{2}{*}{$-$} & \multirow{2}{*}{$-$}\\
			& N (cm$^{-2}$)& $(8.5\pm 2.4)\times 10^{10}$ &	 &  &  & \\ \midrule
			DCN & N (cm$^{-2}$)& $(2.5\pm 0.7)\times 10^{12}$ &	$(5.0\pm 1.1)\times 10^{12}$ & $(2.3\pm 0.5)\times 10^{12}$ & $(3.5\pm 0.7)\times 10^{12}$ & $(3.5\pm 0.6)\times 10^{12}$ \\ \midrule
			\multirow{2}{*}{N$_{2}$H$^{+}$} & N$_{\rm upp}$ (cm$^{-2}$) $^{\bm{(b)}}$ & $(1.2\pm 0.3)\times 10^{13}$ &	$(1.0\pm 0.2)\times 10^{13}$ & $(6.0\pm 1.4)\times 10^{12}$ & $(6.0\pm 1.2)\times 10^{12}$ & $(5.0\pm 0.9)\times 10^{12}$ \\
			& N$_{\rm low}$ (cm$^{-2}$) $^{\bm{(c)}}$ & $(8.0\pm 2.0)\times 10^{12}$ &	$(5.0\pm 1.0)\times 10^{12}$ & $(3.5\pm 0.8)\times 10^{12}$ & $(4.2\pm 0.8)\times 10^{12}$ & $(3.0\pm 0.6)\times 10^{12}$ \\ \midrule
			N$_{2}$D$^{+}$ & N (cm$^{-2}$)& $(1.8\pm 0.5)\times 10^{12}$ &	$(1.5\pm 0.3)\times 10^{12}$ & $<3.5\times 10^{11}$ & $<3.0\times 10^{11}$ & $<3.0\times 10^{11}$ \\ \midrule
			\multicolumn{2}{l}{N(DNC)/N(DCN)} & $2.04\pm 0.82$ & $0.64\pm 0.20$ & $0.87\pm 0.29$ & $0.80\pm 0.22$ & $0.71\pm 0.18$ \\
			\midrule
			\multicolumn{2}{l}{N(N$_{2}$D$^{+}$)/N(N$_{2}$H$^{+}$)$_{\rm upp}$} & $0.15\pm 0.06$ & $0.15\pm 0.04$ & $< 0.06$ & $< 0.05$ & $< 0.06$ \\
			\midrule
			\multicolumn{2}{l}{N(N$_{2}$D$^{+}$)/N(N$_{2}$H$^{+}$)$_{\rm low}$} & ${0.23\pm 0.08}$ & ${0.30\pm 0.08}$ & ${< 0.10}$ & ${< 0.07}$ & ${< 0.10}$\\
			\midrule
			\multicolumn{2}{l}{N(N$_{2}$H$^{+}$)$_{\rm upp}$/N(CS)} & $0.09\pm 0.05$ & $0.08\pm 0.06$ & $0.02\pm 0.01$ & $0.04\pm 0.03$ & $0.02\pm 0.01$ \\
    		\bottomrule
	\end{tabular}}
	\flushleft
			{\small
			\ \ \ \ $^{\bm{(a)}}$ We assume the kinetic temperature $T_{\rm k}$ and the hydrogen nuclei number densities $n_{\rm H}$ obtained from CS and DNC multi-transition fitting (in bold) to derive the DCN, N$_2$H$^+$, and N$_2$D$^+$ column densities.\\
			\ \ \ \ $^{\bm{(b)}}$ Upper bound value for the column density obtained assuming the T$_k$ and n(H$_2$) values in bold and fitting the opacity of the main component and the brightness of the weakest hyperfine component as explained in Section 4.1.\\
			\ \ \ \ $^{\bm{(c)}}$ Lower bound value for the column density obtained in the low excitation temperature scenario discussed in Section 6.\\
			}
	\label{tab:col_dens}
\end{table*}

As discussed previously, deuterated molecules are suitable for tracing dense and cold regions inside dark molecular clouds. Given the necessity to characterize these regions for future modeling, we estimated the temperature, density, and column densities across our targets. We derived the column density of CS by fitting the emission of the $^{13}$CS $2\rightarrow1$, C$^{34}$S $3\rightarrow2$, and CS $3\rightarrow2$ lines following the procedure already described in \citet{Fuente2019} and \citet{NavarroAlmaida2020}. Briefly, this method explores a parameter space consisting in kinetic temperatures (T$_k$), molecular hydrogen densities ($n_{\rm H_{2}}$), and column densities of the species following the Markov Chain Monte Carlo (MCMC) methodology with a Bayesian inference approach, described in \citet{Riviere2019}. To be consistent, we smoothed the maps of the C$^{34}$S $3\rightarrow2$ and CS $3\rightarrow2$ lines to the angular resolution of those of the $^{13}$CS $2\rightarrow1$ line. This method uses the radiative transfer code {\scshape Radex} \citep{vanderTak2007} to reproduce all the line temperatures. Since our observations include different isotopologs, for this method to work we need to reduce the degrees of freedom that these different species introduce. To do so we assume the isotopic ratios N(C$^{34}$S)/N($^{13}$CS) = 8/3 and N(CS)/N($^{13}$CS) = 60 \citep{Gratier2016}. For consistency in the radiative transfer calculations presented in this paper, we only considered H$_{2}$ as the collisional partner for CS and isotopologs, with the collisional coefficients described in \citet{DenisAlpizar2013}. The effect of ignoring collisions with helium would translate in densities that are $\sim$20\% higher (He/H$_2$=0.20), which is within the uncertainties of our calculations. The resulting kinetic temperatures, hydrogen nuclei densities, and CS column densities are shown in Table \ref{tab:col_dens}. The synthetic $^{13}$CS $2\rightarrow1$, C$^{34}$S $3\rightarrow2$, and CS $3\rightarrow2$ line profiles resulting from these parameters are shown in Figure~\ref{fig:csRadex}, together with the observations.

Since our setup only considers up to two lines of DCN and DNC, it is necessary to assume the kinetic temperature at each position to derive reliable hydrogen nuclei densities and molecular column densities. We assumed the kinetic temperatures at these positions as those given by the multi-transition analysis of the CS (and isotopologs) lines, as they are in agreement with those calculated in \citet{Fuente2019} and \cite{NavarroAlmaida2020}. The observation of two lines of DNC and DN$^{13}$C allows for an independent estimate of the hydrogen nuclei density. We derived the hydrogen nuclei density and DNC (DN$^{13}$C) column density at each region using the radiative transfer code {\scshape Radex} and the HNC collisional coefficients by \citet{Dumouchel2010}. As mentioned in the previous section, the DNC $1\rightarrow0$ line is expected to be optically thick towards the position TMC~1-C. In this position, the multi-transition study of the rarer isotopolog DN$^{13}$C provides more accurate values. To calculate the DNC column density, it is assumed that DNC/DN$^{13}$C $\sim 60$, a value commonly accepted for TMC1. We recall however that the DNC/DN$^{13}$C ratio itself is uncertain to a factor of 1.3 \citep{Liszt2012, Roueff2015}. The densities derived from CS are a factor $\sim$ 3 to $\sim 6$ lower than those derived from DNC and DN$^{13}$C (see Table \ref{tab:col_dens}). This discrepancy could be due  to different radial distributions with the deuterated compounds coming mainly from the densest regions where CS is expected to be depleted. Since DCN and DNC are expected to trace the innermost regions of our targets, we take the densities we derived from these molecules as more appropriate to describe the cores. The volume densities in TMC~1-CP and TMC~1-C differ by a factor of 2. This difference is within the uncertainties of our calculations, and we cannot conclude that TMC~1-C is more evolved than TMC~1-CP only on the basis of the density estimations. Interestingly, we do not find any significant difference in volume density between the different positions within the same core, indicating a uniform density distribution for R $<6000$ au.  We recall that our beam is $\sim$5000~au at frequencies of the deuterated species lines.
 
We adopted the densities derived from DNC and estimate the DCN column density by fitting the weakest hyperfine component of the J=$1\rightarrow0$ line. This is a good approximation as long as the opacity of this line is  moderate. Assuming that the excitation temperature is the same for all the hyperfine components, we can estimate the opacity of the DCN $1\rightarrow0$ line using the HFS method of GILDAS-CLASS (see Table \ref{tab:hfs}). In the case of TMC~1-C, the opacities are so high that we might be still underestimating the DCN column density. The opacity of the main lines is $\tau_m\gtrsim 4$, implying that the opacity of the weakest line would be $\sim$1. Thus, our fitting could underestimate the DCN column density by a factor of $\sim$2. Therefore, we consider the derived DCN column density towards TMC~1-C as a lower limit to the real value. Synthetic spectra are created using the values in Table~\ref{tab:col_dens}, which are compared with observations in Figure~\ref{fig:spectra}.

\subsection{N$_2$H$^+$ and N$_2$D$^+$}

In the cases of N$_2$H$^+$ and N$_2$D$^+$, we observed one transition and therefore we cannot carry out a multi-transition study. We can however take advantage of the hyperfine splitting to quantify the opacities. As explained above, we can use the GILDAS-CLASS HFS method to estimate the line opacities and calculate accurate column densities (Figure~\ref{fig:spectra}). This method computes the line profiles of a species and the opacity of the main component given the relative velocities and intensities of the components to the most intense component. In our analysis the relative intensities and velocities are fixed to laboratory values. This method assumes a Gaussian velocity distribution and the same line width and excitation temperature for each component. However, the N$_2$H$^+$ $1\rightarrow 0$ spectra towards TMC~1-C and TMC~1-CP cannot be reproduced using this technique, which suggests that we need more sophisticated modeling with at least two gas layers to account for the N$_2$H$^+$ observations. Nevertheless, we can obtain a first estimate of the N$_2$H$^+$ column density by exploring two different scenarios that can be considered as limit cases. First, assuming the values of $T_{\rm k}$ and $n_{\rm H}$ derived from the CS and DNC fitting, we estimate the N$_2$H$^+$ column density in such a way that the brightness of the weakest hyperfine component matches the observations. We fit the intensity of the weakest hyperfine component N$_2$H$^{+}$ $J_{F_{1},F}=1_{1,0}\rightarrow 0_{1,1}$ because it is the least affected by opacity effects. This can be considered as a high excitation temperature scenario, where the N$_2$H$^{+}$ column density is an upper bound value (see Table \ref{tab:col_dens}). The resulting spectrum of N$_2$H$^{+}$ $1\rightarrow 0$ is plotted in Figure~\ref{fig:spectra}, and overestimates the temperatures of the rest of components. For these to be fitted, a second scenario of low excitation temperatures should be considered. The appropriate excitation temperatures are obtained solving the radiative transfer equation for spectral lines with {\scshape Radex}. These low temperatures, of around $T_{\rm ex}\sim 4$ K in TMC 1-C and $T_{\rm ex}\sim 5$ K in TMC 1-CP, correspond to environments with either low kinetic temperatures ($\sim 5$ K in TMC 1-C and $\sim 6-7$ K in TMC 1-CP) or relative low densities ($n_{\rm H}\sim 10^{4}$ cm$^{-3}$). Thus, the N$_2$H$^+$ emission would be linked to dense regions at extremely low temperatures, lower than the temperatures derived from our observations (see Table \ref{tab:col_dens}). Alternatively, the emission could originate in an optically thick envelope at low density. Proceeding as before, we obtain a lower bound for the N$_2$H$^+$ column density such that both the brightness temperature and the opacity of the main component match the observations. In this case, the intensity of the weakest hyperfine line is underestimated, especially for TMC~1-C. The column densities obtained in the two scenarios are found to agree within a factor of two and are summarized in Table \ref{tab:col_dens}. To better interpret our results we must take into account that self-absorption features are present in the spectra of other transitions in our setup, such as DCN $1\rightarrow 0$. These features point towards the presence of an absorbing envelope around TMC 1-C. This possibility has been already discussed in \citet{Fuente2019}, and its effect on the observations is explored in Section 10. We propose that the low brightness temperatures of the optically thick N$_2$H$^+$ hyperfine lines is the consequence of self-absorption by the optically thick low-density envelope. Self-absorption has less affect on the optically thin emission of the weakest hyperfine line. If our interpretation is correct, the high excitation temperature scenario is more representative of the TMC 1-C and TMC 1-CP environments. Thus, we adopt the upper bound values for the N$_2$H$^+$ column density as the ones that best describe the observations with an uncertainty of a factor of 2. 

In addition to N$_2$H$^{+}$ $1\rightarrow 0$, we also include the N$_2$D$^{+}$ $1\rightarrow 0$ line in our setup. This time, however, the spectra have the appropriate line intensity ratios. As before, we obtain the N$_2$D$^{+}$ column density in TMC1-C by matching the opacity of the main component and the brightness of the hyperfine components, assuming the values of T$_k$ and n$_H$ derived from the CS and DNC fitting. We obtain  N(N$_2$D$^+$)/N(N$_2$H$^+$) $\sim 0.15 \pm 0.06$ in TMC~1-C assuming the high value of N(N$_2$H$^+$) (see Table \ref{tab:col_dens}). We did not detect N$_2$D$^+$ towards TMC~1-CP with N(N$_2$D$^+$)/N(N$_2$H$^+$) $<0.06$. \citet{Crapsi2005} analyzed the N$_{2}$H$^{+}$ and N$_{2}$D$^{+}$ abundances in a large sample of starless cores. They calculated the N$_{2}$H$^{+}$ deuteration fraction in TMC~1-C, obtaining a N$_{2}$D$^{+}$/N$_{2}$H$^{+}$ ratio of $0.07\pm 0.02$, consistent with our results. Their estimation in their position named TMC 1, which is near the cyanopolyyne peak, yields a lower N$_{2}$D$^{+}$/N$_{2}$H$^{+}$ ratio $0.04\pm 0.01$, compatible with our estimated upper bound in the position TMC~1-CP. 

As noted before, the N$_2$H$^+$ column density and the N(N$_2$D$^+$)/N(N$_2$H$^+$) ratio are considered evolutionary diagnostics for starless cores. Regardless of the different scenarios proposed to estimate the N$_2$H$^+$ column density, N(N$_2$H$^+$) is higher in TMC~1-C than in TMC~1-CP, suggesting that the former is in a more evolved stage, although the difference is within the uncertainties. Similarly, the higher value of N(N$_2$D$^+$)/N(N$_2$H$^+$) estimated towards TMC~1-C, taking into account all possible values of N(N$_2$H$^+$), does confirm the more evolved stage of this core. 

\begin{table*}
	\centering
	\caption{Model parameters and initial fractional abundances with respect to total hydrogen nuclei.}
	\resizebox{0.85\textwidth}{!}{
	\begin{tabular}{lccccccc}
		\toprule
		Parameter & & Model A & Model B & Model C & Model D \\\midrule
				T$_{k}$ (K) & Temperature & \multicolumn{4}{c}{$9$} & & \\
				A$_{v}$ (mag) & Visual extinction & \multicolumn{4}{c}{20 mag} & & \\ 
				HD & Hydrogen deuteride abundance & \multicolumn{4}{c}{$1.60\times 10^{-5}$} & & \\
				He & Helium abundance & \multicolumn{4}{c}{$9.00\times 10^{-2}$} & & \\
				N & Nitrogen abundance & \multicolumn{4}{c}{$6.20\times 10^{-5}$} & & \\
				O & Oxygen abundance & \multicolumn{4}{c}{$2.40\times 10^{-4}$} & & \\
				C$^+$ & Carbon abundance & \multicolumn{4}{c}{$1.70\times 10^{-4}$} & & \\
				S$^+$ & Sulfur abundance & \multicolumn{4}{c}{$1.50\times 10^{-5}$}  & & \\
				Si$^+$ & Silicon abundance & \multicolumn{4}{c}{$8.00\times 10^{-9}$} & & \\
				Fe$^+$ & Iron abundance & \multicolumn{4}{c}{$3.00\times 10^{-9}$} & & \\
				Na$^+$ & Sodium abundance & \multicolumn{4}{c}{$2.00\times 10^{-9}$} & & \\
				Mg$^+$ & Magnesium abundance & \multicolumn{4}{c}{$7.00\times 10^{-9}$} & & \\
				P$^+$ & Phosphorus abundance & \multicolumn{4}{c}{$2.00\times 10^{-10}$} & & \\
				Cl$^+$ & Chlorine abundance & \multicolumn{4}{c}{$1.00\times 10^{-9}$} & & \\
				F & Fluorine abundance & \multicolumn{4}{c}{$6.68\times 10^{-9}$} & & \\
				$\chi_{\rm UV}$ & UV field strength (Draine units) & \multicolumn{4}{c}{5} & & \\ \midrule
				o-H$_{2}$/p-H$_{2}$ & Initial ortho-to-para ratio of H$_{2}$ & 3 & 3 & $10^{-3}$ & $10^{-3}$\\
				$\zeta$(H$_{2}$) (s$^{-1}$) & Cosmic-ray ionization rate & $1.3\times 10^{-17}$ & $1\times 10^{-16}$ & $1.3\times 10^{-17}$ & $1\times 10^{-16}$ \\
			    \bottomrule
	\end{tabular}}
	\label{tab:model_param}
\end{table*}

\section{Physical and chemical modeling of starless cores}
\label{sec:model}

The physical and chemical diagnostics used in the previous section (i.e., the volume density at the center of the core), the N$_2$H$^+$ column density, and the N$_2$D$^+$/N$_2$H$^+$ column density ratio indicate that TMC~1-C is in a later evolutionary stage. We test and develop this idea with the help of chemical models in the following sections.

\subsection{Chemical network and general considerations}

We modeled the chemistry of  TMC~1-CP and TMC~1-C using the \textsc{Nautilus} gas-grain chemical code \citep{Ruaud2016}. \textsc{Nautilus} is a three-phase model in which gas, grain surface, and grain mantle phases, and their interactions, are considered. \textsc{Nautilus} solves the kinetic equations for both the gas-phase and the surface of interstellar dust grains and computes the evolution with time of chemical abundances for a given physical structure. We use the chemical network presented by \citet{Majumdar2017}, which considers multiple deuterated molecules and includes the spin chemistry in \textsc{Nautilus}. In all models, we adopted the initial abundances and the ambient UV field strength $\chi_{UV}$=5 (in Draine units), determined by \citet{Fuente2019}, and shown in Table~\ref{tab:model_param}. In the next section we use zero-dimensional (0D) chemical models to explore the influence of each parameter in the deuterium chemistry. In Section~\ref{sec:pseudo} we carry out one-dimensional (1D) pseudo-dynamical calculations to constrain the evolutionary stage of these cores.

Modeling CS observations with chemical models is challenging due to well-known uncertainties in the chemistry of CS (see, e.g., \citealp{Vidal2017, Laas2019, NavarroAlmaida2020, Roncero2020}). Based on H$_2$S, SO, and CS observations, \citet{NavarroAlmaida2020} conclude that sulfur elemental abundance in TMC~1 is equal to the solar value within a factor of 10. Recent calculations by \citet{Roncero2020} concluded that CS abundance in TMC~1 is overestimated when assuming the solar sulfur abundance; in order to reproduce observations of CO, CS, and HCS$^{+}$ consistently, a sulfur depletion by a factor of 20 is needed. In this section we assume undepleted sulfur abundances as in \citet{NavarroAlmaida2020}, and we discuss the effect of possible sulfur depletion in Sect~\ref{sec:mock}. Taking into account the uncertainties in the chemical modeling of CS, we base our discussion of the best model mainly on the abundances of N$_2$H$^+$ and the deuterated species N$_2$D$^+$, DCN, and DNC. The abundances of these molecules remain essentially unaffected by a variation of the sulfur elemental abundance of a factor of 20.

\begin{figure*}
		\includegraphics[width=\textwidth,keepaspectratio]{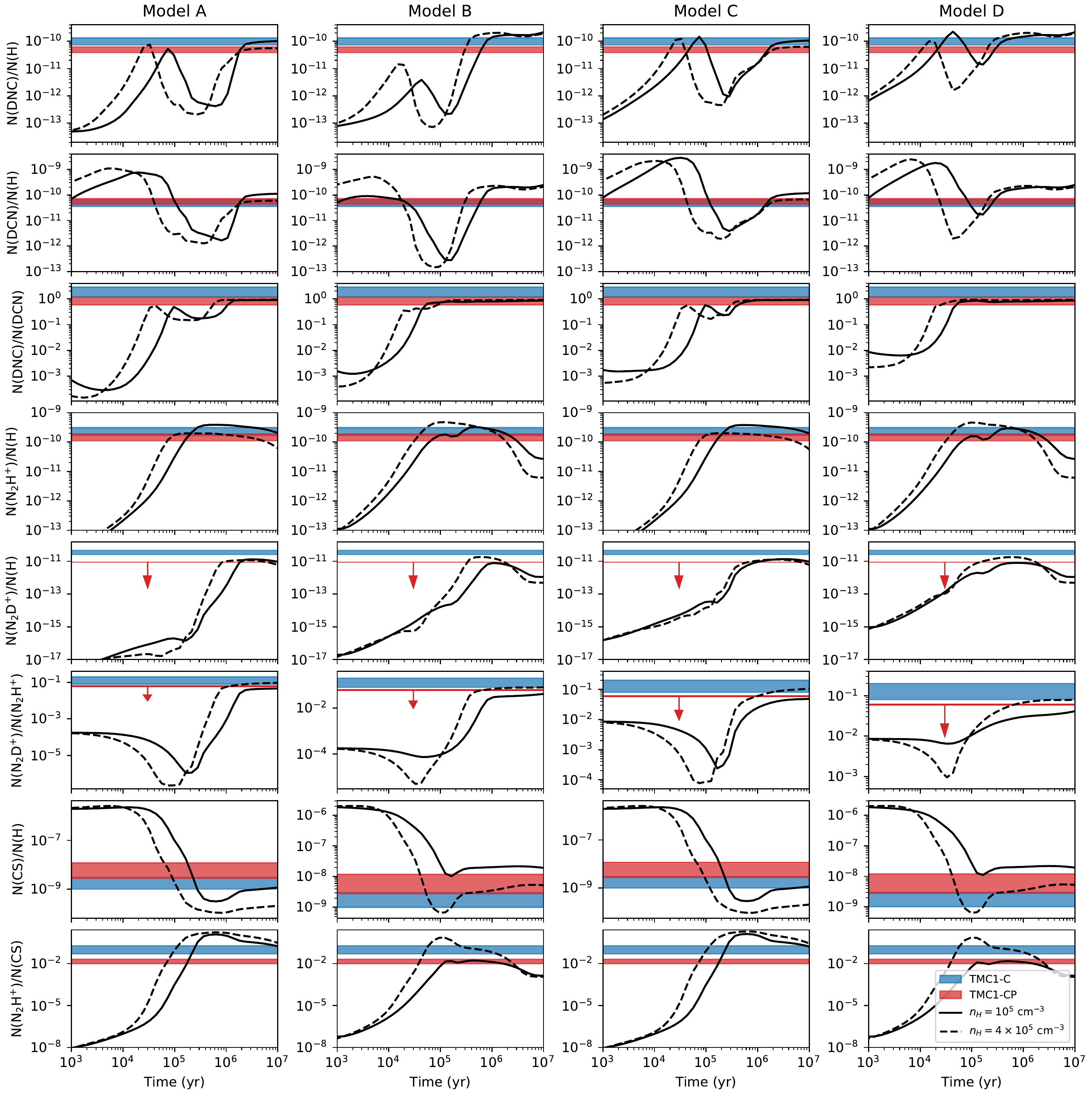}
		\caption{Abundances and abundance ratios of the different molecules, as predicted by the models described in Table \ref{tab:model_param}. The predicted values are shown in solid (model with $n_{\rm H}=10^{5}$ cm$^{-3}$) and dashed lines (model with $n_{\rm H}=4\times 10^{5}$ cm$^{-3}$). The blue and red areas show the observed abundances in TMC~1-C and TMC~1-CP, respectively. The arrows indicate upper bound values.}
		\label{fig:deut_time_ev}
\end{figure*}

\begin{figure*}
		\includegraphics[width=\textwidth,keepaspectratio]{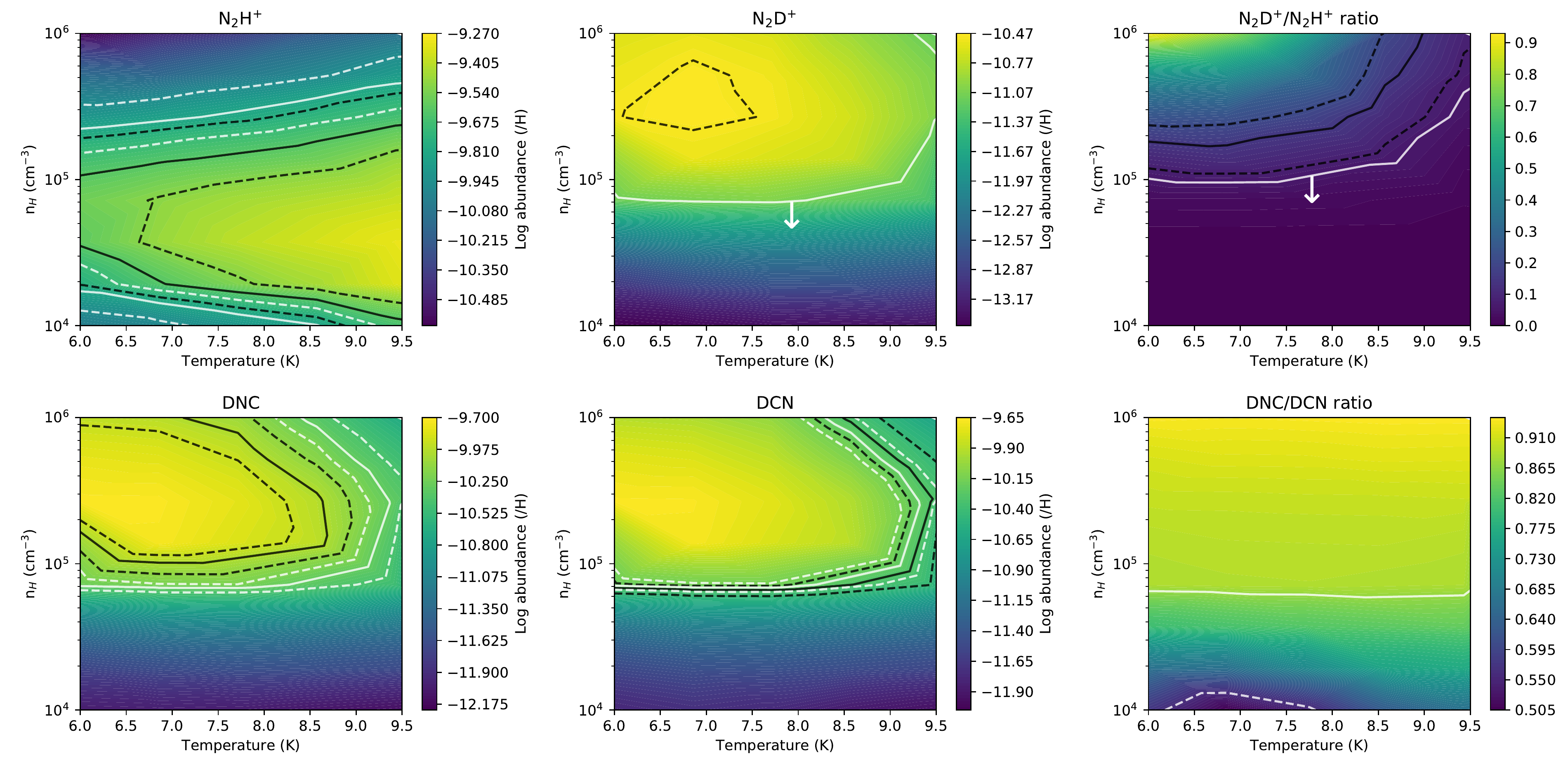}
		\caption{Grids of \textsc{Nautilus} models with varying temperature and hydrogen nuclei density. Each plot shows the contour levels of the N$_{2}$H$^{+}$ (top left), N$_{2}$D$^{+}$ (top middle), DNC (bottom left), and DCN (bottom middle) abundances; the N$_{2}$H$^{+}$ deuterium fraction (top right); and the DNC/DCN ratio (bottom right) at time $t\approx 10^{6}$ yr. Additionally, we show the contours that correspond to the abundances derived from the N$_{2}$H$^{+}$, N$_{2}$D$^{+}$, DNC, and DCN column densities in TMC~1-C (black) and TMC~1-CP (white), their uncertainties (dashed lines), and the upper bound values if present (white arrows). The model does not reproduce our estimated DNC/DCN ratio towards TMC~1-C, and thus the black contours are missing from that panel.}
		\label{fig:grid}
\end{figure*}

\subsection{Zero-dimensional models}
Chemistry has become an essential tool in determining the physical conditions along starless cores. However, pseudo-time dependent and time-dependent chemical models show that chemical abundances depend on parameters that are usually not well constrained, such as the cosmic-ray ionization rate $\zeta({\rm H}_2)$ and the ortho-to-para ratio (OPR) of H$_2$ \citep{Roueff2007, Roueff2015, Majumdar2017, Sipila2018}. In the following, we use 0D models to explore the parameter space and determine suitable parameters for our pseudo-dynamical calculations.

Table \ref{tab:model_param} summarizes the grid of models considered in our exploratory study. In dense cores where most of the hydrogen is expected to be in molecular form, cosmic rays ionize H$_{2}$ to yield H${_{3}}^{+}$ ions \citep{Herbst1973, Dalgarno2006}, enhancing the deuteration through Eq. \eqref{eq:deut}. \citet{Fuente2019} estimated $\zeta({\rm H}_2)$= (0.5-1.8) $\times$ 10$^{-16}$ s$^{-1}$ in TMC~1 based on the chemical modeling of the moderate density gas ($n_{\rm H}$$\sim$2$\times$10$^4$ cm$^{-3}$). This value, however, is not representative of the central regions of TMC~1-CP and TMC~1-C where we estimate densities one order of magnitude higher and the visual extinction is higher \citep{Padovani2013, Galli2015, Ivlev2015, Padovani2020}. Consequently, in our grid of models we considered two values of $\zeta({\rm H}_2)$, the average interstellar value 1.3$\times$10$^{-17}$ s$^{-1}$ and 1$\times$10$^{-16}$ s$^{-1}$. The reverse of Eq. \eqref{eq:deut} is more efficient when involving the ortho-H$_{2}$ species than the para-H$_{2}$ analogs \citep[see, e.g.,][]{Gerlich2002, Flower2006}. Thus, the ortho-H$_{2}$ abundance sets the availability of H$_{2}$D$^{+}$, the main parent species of deuterated molecules, at low temperatures. Although ortho- and para-H$_2$ are formed on the surfaces of interstellar grains with a statistical ratio of 3:1 \citep{Watanabe2010}, under the physical conditions prevailing in dark clouds (T$_k<10$ K) the OPR decreases to values $\sim$10$^{-3}$ after 0.1 Myr \citep{Majumdar2017}. In the case of a static pre-phase before collapse, the initial OPR would be closer to  $\sim$10$^{-3}$ than to 3. In order to explore the impact of this parameter in our modeling, we considered two values of initial OPR, 3 and 10$^{-3}$ in the 0D calculations.

Using the models in Table \ref{tab:model_param}, we ran \textsc{Nautilus} to predict the chemical evolution of the DNC, DCN, N$_{2}$H$^{+}$, N$_{2}$D$^{+}$ for two densities, $n_{\rm H} = 10^{5}$ cm$^{-3}$ and $n_{\rm H} = 4\times 10^{5}$ cm$^{-3}$,  which are the values estimated for TMC~1-CP and TMC~1-C. The predicted abundances for each model  up to $t = 10$ Myr are shown in Figure~\ref{fig:deut_time_ev}. In addition, we indicate the abundances observed towards TMC~1-C and TMC~1-CP with colored horizontal bands. The comparison of models A and B (C and D) inform us about the influence of  $\zeta({\rm H}_2)$ on the molecular abundances. Once the chemistry has reached a steady state, the abundances of the deuterated molecules DCN and DNC are enhanced in Models B and D, relative to A and C. However, the reverse stands for the ionized species N$_{2}$H$^{+}$ and N$_{2}$D$^{+}$ that are rapidly destroyed via dissociative recombination with  electrons \citep{Dalgarno2006}. \citet{NavarroAlmaida2020} estimated $t \sim 1$ Myr for these cores based on the abundances of the S-bearing species. Assuming that the chemical age of TMC~1-CP and TMC~1-C is $\sim$ 1 Myr, the abundances of DCN, DNC, N$_2$H$^+$, and N$_2$D$^+$ are better reproduced with $\zeta({\rm H}_2)=1.3\times 10^{-17}$ s$^{-1}$, which is a factor of $\sim 5-10$ lower than that estimated by \citet{Fuente2019} in the TMC~1 translucent filament. Several works have investigated the attenuation of $\zeta({\rm H}_2)$ with the H$_2$ column density into the cloud \citep{Galli2015, Ivlev2015, Padovani2020}. \citet{Galli2015}, based on observations ranging from diffuse clouds to molecular cloud cores, concluded that the cosmic-ray ionization rate decreases towards higher column density objects.  \citet{Neufeld2017} also established a dependence of the cosmic-ray ionization rate with the visual extinction, allowing a change of a factor of 5$-$10 of this rate in the  extinction gradients found in our targets. 

We selected a high ortho-to-para ratio of H$_{2}$ of 3 (models A and B) and a low value of $\sim 10^{-3}$ (models C and D) to estimate the influence of this parameter in the predicted chemical abundances. In Figure~\ref{fig:deut_time_ev} we see that the abundances of the deuterated molecules in all models reach similar values beyond 0.1 Myr. This result was already found by \citet{Majumdar2017} who concluded that the OPR decreases to values $<$10$^{-3}$ for $t > 0.1$ Myr, regardless of its initial value. For times $t < 0.1$ Myr there is an enhancement in the abundance of deuterated molecules of the models C and D compared to models A and B. Our cores are expected to be older than 0.1~Myr \citep{NavarroAlmaida2020, Roncero2020}; therefore, an OPR=10$^{-3}$ is adopted in the following calculations.

One crucial point for our study is to check the ability of the observed deuterated species to constrain the chemical age. Figure~\ref{fig:deut_time_ev} shows that the abundances of DCN, DNC, and  N$_2$D$^+$, as well as the DNC/DCN and N$_2$D$^+$/N$_2$H$^+$ ratios, are strongly dependent on the time evolution. Interestingly, the DNC/DCN ratio reaches its equilibrium value at earlier times than N$_2$D$^+$/N$_2$H$^+$. The combination of the two ratios therefore provides a useful tool for constraining the value of $\zeta({\rm H}_2)$  and the chemical age of the cores with $t \sim 0.1- 1$~Myr.

Once the values of key parameters such as the cosmic-ray ionization rate and the initial ortho-to-para ratio of H$_{2}$ are constrained, we can proceed to investigate the behavior of the N$_{2}$H$^{+}$, N$_{2}$D$^{+}$, DNC, and DCN with variations in density and temperature. We carry out an exploration of the $n_{\rm H}$-T parameter space considering 0D models with varying densities and temperatures, a cosmic-ray ionization rate of $\zeta({\rm H}_2)=1.3\times 10^{-17}$ s$^{-1}$, an OPR of 10$^{-3}$, and a chemical age of $\sim 1$~Myr. In Figure~\ref{fig:grid}, we show the abundances of N$_2$H$^+$, N$_2$D$^+$, DCN, DNC, as well as the N$_2$D$^+$/N$_2$H$^+$ and DNC/DCN abundance ratios, as a function of the gas kinetic temperature and density. The abundances of N$_2$D$^+$, DCN, and DNC depend on the gas kinetic temperature. In particular, the abundances derived towards TMC~1-CP and TMC~1-C are consistent with model predictions in the range of T$_k$$\sim$6$-$9~K, in agreement with the temperatures derived in our CS multi-transition study. Interestingly, the abundance of N$_2$H$^+$ and the DNC/DCN ratio have a very weak dependence with the temperature. The abundances of the deuterated isomers DCN and DNC, and mainly the DNC/DCN ratio, increase with the number density of total hydrogen nuclei. According to our modeling, these abundances are expected for gas densities, $n_{\rm H}\sim 2\times 10^5$ cm$^{-3}$, a factor of 2 lower than those calculated from our multi-transition study of DCN and DNC. The deuterated compound N$_2$D$^+$ has only been detected towards TMC~1-C, and its abundance is consistent with a density, $n_{\rm H}\sim 5\times 10^5$ cm$^{-3}$, compatible with our density estimate using the DN$^{13}$C isotopolog. N$_2$D$^+$ was not detected in TMC~1-CP, with an upper limit to the N$_2$D$^+$/N$_2$H$^+$ ratio that would correspond to $n_{\rm H}$ $<$ 2$\times$10$^5$ cm$^{-3}$, slightly lower than our density estimate.
 
\subsection{One-dimensional pseudo-dynamical modeling}
\label{sec:pseudo}
 
Several works have simulated the chemical evolution of a core undergoing collapse with different hydrodynamical models and with different treatments for the coupling between dynamics and chemistry \citep{Rawlings1992c, Bergin1997, Aikawa2001, Aikawa2005, Sipila2018, PriestleyViti2018}. In some cases the simulations were continued until the formation of the Larson core, describing the evolution from the pre-stellar to the proto-stellar phase \citep{Aikawa2008, Hincelin2016}. The large chemical network used in our modeling, with 7700 reactions on grain surfaces linked with the 111000 reactions in gas phase, precludes its usage in chemo-hydrodynamical calculations. Instead, we used the parametric expression derived by \citet{PriestleyViti2018} to reproduce the results of 1D simulations of collapsing cores. In particular, they adopted the gravitational collapse of a spherical cloud described by \citet{Aikawa2005}. In this model, the collapse is induced by a density increase by a factor of either 1.1 (BES1 model) or 4 (BES4 model) to produce a gravitational free-fall (t$\sim$1 Myr) in model BES1 or a rapid collapse induced by cloud compression (t$\sim$0.1 Myr) in model BES4. A similar study was carried out by \citet{Fiedler1993}, but adding ambipolar diffusion to a magnetically supported core (AD model), which translated into a larger timescale t$\sim$10 Myr. These models were parameterized as a sequence of progressively dense Plummer-like spheres whose peak densities depend on the time elapsed since the onset of the collapse \citep{PriestleyViti2018}. Plummer-like profiles have been successfully used to model the density of the TMC~1 cores \citep[see, e.g.,][]{Schnee2010, NavarroAlmaida2020} which makes it plausible to assume this model.

\begin{table*}
	\centering
	\caption{Model parameters, including the collapse time $t$, the central hydrogen nuclei number density $n_{0}$, the flat radius $r_{0}$, and the asymptotic power index $a$.}
	\begin{tabular}{lcccc}
		\toprule
		Model & $n_{0}=5\times 10^{4}$ cm$^{-3}$ & $n_{0}=1\times 10^{5}$ cm$^{-3}$ &  $n_{0}=4\times 10^{5}$ cm$^{-3}$ & $n_{0}=1\times 10^{6}$ cm$^{-3}$ \\ \midrule
				\multirow{3}{*}{BES1} & $t = 5.73\times 10^{5}$ yr & $t = 8.29\times 10^{5}$ yr & $t = 1.06\times 10^{6}$ yr & $t = 1.12\times 10^{6}$ yr\\
				& $r = 9.58\times 10^{3}$ au & $r = 6.96\times 10^{3}$ au & $r = 3.67\times 10^{3}$ au & $r = 2.41\times 10^{3}$ au\\
				& $a = 2.40$ & $a = 2.40$ & $a = 2.40$ & $a = 2.40$\\ \midrule
				\multirow{3}{*}{BES4} &  & $t = 3.18\times 10^{4}$ yr & $t = 1.28\times 10^{5}$ yr & $t = 1.56\times 10^{5}$ yr\\
				& $-$ & $r = 1.23\times 10^{4}$ au & $r = 5.59\times 10^{3}$ au & $r = 3.31\times 10^{3}$ au\\
				&  & $a=2.26$ & $a=2.04$ & $a=2.00$\\ \midrule
				\multirow{3}{*}{AD} & $t = 1.57\times 10^{7}$ yr & $t = 1.58\times 10^{7}$ yr & $t = 1.60\times 10^{7}$ yr & $t = 1.61\times 10^{7}$ yr \\
				& $r = 1.89\times 10^{4}$ au & $r = 1.35\times 10^{4}$ au & $r = 6.88\times 10^{3}$ & $r = 4.13\times 10^{3}$ au\\
				& $a = 2.34$ & $a = 2.30$ & $a = 2.25$ & $a = 2.23$\\
    		\bottomrule
	\end{tabular}
	\label{tab:models}
\end{table*}

In the parameterization proposed by \citet{PriestleyViti2018}, the density in the core center and the radial density profile is determined by the time since the onset of collapse. In Table~\ref{tab:models} we show the models considered to compare with the molecular abundances observed in TMC~1-C and TMC~1-CP. Given the central density ($n_0$) and the type of collapse (BES1, BES4, AD), the density profile is described by the parameters $a$ and r$_0$ by the expression

\begin{equation}
	n (r) = n_0/(1 + r / r_0)^a,
\end{equation}
where $n_{0}$ is the central density, $r_{0}$ is the flat radius, and $a$ is the asymptotic power index. The gas temperature across the core is assumed equal to the dust temperature that is described by

\begin{equation}
	T_d (r) = T_0 (1 + r / r_1),
\end{equation}
with $T_{0}=8.5$ K and $r_{1}=3.66\times 10^{4}$ au, representative central temperature and size of the cores \citep{NavarroAlmaida2020}. These density and temperature profiles are used as the physical structure for our 1D chemical modeling. Our simulations start with the gas in atomic form and the chemistry is allowed to evolve during the collapse time shown in Table~\ref{tab:models}. Following our findings, we adopt  $\zeta({\rm H}_2)=1.3\times 10^{-17}$ s$^{-1}$ and OPR $=10^{-3}$ in our 1D calculations. The initial elemental abundances are the same as in \citet{NavarroAlmaida2020}. As output of our simulations we obtain the molecular abundances as a function of the radius for the eight models shown in Table~\ref{tab:models}. In order to compare models with observations, we assume a spherical geometry and calculate weight-average column densities along the line of sight as

\begin{equation*}
	{\rm [X]_{\rm ac}}(r) = \frac{\sum_{i}\Delta l_{i}\left(n_{\rm H}(s_{i}){\rm [X]}(s_{i})+n_{\rm H}(s_{i+1}){\rm [X]}(s_{i+1})\right)}{\sum_{j}\Delta l_{j}\left(n_{\rm H}(s_{j})+n_{\rm H}(s_{j+1})\right)},
\end{equation*}
where $r$ is the impact parameter; $\Delta l_{i}= l_{i+1}-l_{i}$; $s_{i}=\sqrt{r^2+l_{i}^{2}}$; and $l_{i}$ is a discretization of the segment along the line of sight $l_{\rm max} > \dots > l_{i+1} > l_{i} > \dots > 0$, with $l_{\rm max} = \sqrt{r_{\rm max}^{2}-r^{2}}$, and $r_{\rm max}$ the radius of the density profile. The obtained abundance ratios as a function of the impact parameter are shown in Figures.~\ref{fig:bes1}, \ref{fig:bes4}, and \ref{fig:ad}. 

\begin{figure*}
	\centering
	\includegraphics[width=0.9\textwidth,keepaspectratio]{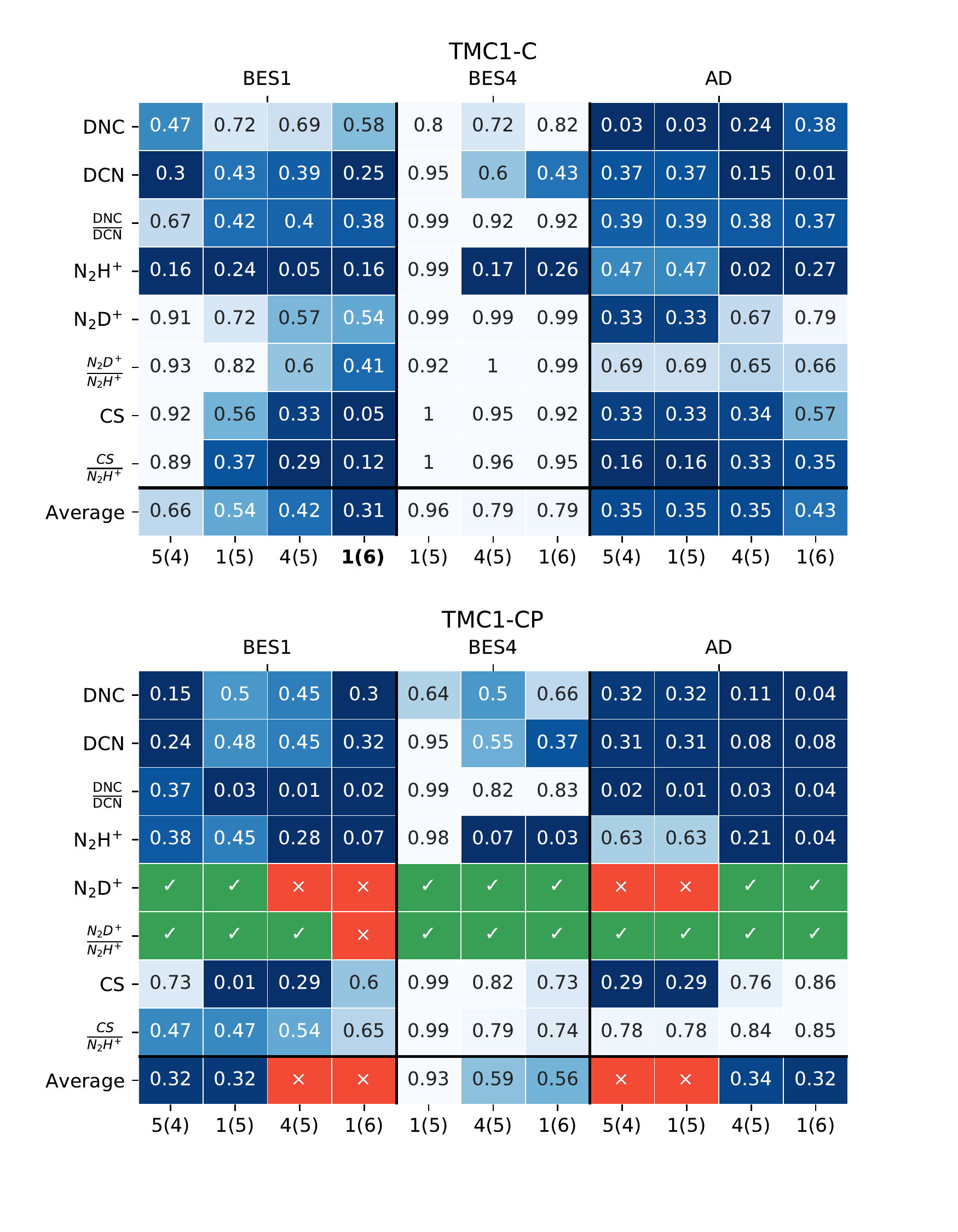}
	\caption{Heat maps of relative differences $\Delta = \frac{|\rm model\ -\ measurement|}{\rm model\ +\ measurement}$ (dark blue means better agreement) and their averages between the predictions of each model (BES1, BES4, and AD, upper x-axis) and the observed abundances and ratios. Each column corresponds to a model with density $n_{\rm H}= a\times 10^{b}$, displayed as $a(b)$. {\itshape Top:} heat map of relative differences in TMC~1-C. The best-fitting model is shown in bold. {\itshape Bottom:} heat map of relative differences in TMC~1-CP. The green and red models are those in agreement and disagreement, respectively, with the observed N$_{2}$D$^{+}$ and N$_{2}$D$^{+}$/N$_{2}$H$^{+}$ upper bounds.}
	\label{fig:heatmap}
\end{figure*}

To assess how well models reproduce the observations, we compute the relative difference $\Delta$ between the modeled weight-average column densities and those observed towards TMC~1-C and TMC~1-CP. For convenience, we define this quantity to be in the $[0,1]$ interval $\Delta = \frac{|\rm model\ -\ measurement|}{\rm model\ +\ measurement}$ such that differences closer to zero indicate well-fitted empirical data while those closer to one mean a poor agreement between the model and the observations. The values of $\Delta$ for each target are shown in Figure~\ref{fig:heatmap}. The results show the poor performance of the fast collapse model BES4 compared to the BES1 and AD models. This is not surprising since the time since collapse in BES4 are  $<1$ Myr (i.e., the timescale for deuteration). This lack of compatibility favors the scenario of a slower collapse in TMC~1.

For TMC~1-C the best model is found to be BES1 with a density of  $n_{\rm H} \sim 1\times 10^{6}$ cm$^{-3}$, consistent with the results obtained in the 0D modeling. The AD models with $n_{\rm H} > 4\times 10^{5}$ cm$^{-3}$ underestimate the N$_2$D$^+$ abundance. However, we also find a good match with observations for AD with  density of $n_{\rm H}= 1\times 10^{5}$ cm$^{-3}$. Since the density of this model is lower than that estimated from our observations, we are prone to favor the BES1 model with higher density. Due to the increasing depletion of molecules with density and projection effects along the line of sight, we would expect the density at the center to be higher than the estimated from our multi-transition study. Moreover, \citet{Schnee2010} proposed that the peak density could be as large as $n_{\rm H} \sim 2\times 10^{6}$ cm$^{-3}$ based on the modeling of the dust continuum emission. Furthermore, the evolutionary time corresponding to this model, $t \sim$ 1 Myr (see Table~\ref{tab:models}), is in better agreement with the previous estimation by \citet{Roncero2020} based on CO, CS, and HCS$^+$ observations. In order to test this hypothesis, in Section~\ref{sec:mock} we perform 3D radiative transfer calculations to produce synthetic lines to compare with our observations.

In TMC~1-CP we need to take into account the upper bound values of the N$_{2}$D$^{+}$ abundance and N$_{2}$D$^{+}$/N$_{2}$H$^{+}$ ratio in order to constrain the central density and the chemical age. Assuming the BES1 model, we derive an upper bound to the density of $n_{\rm H} < 4\times 10^{5}$ cm$^{-3}$ and  $t < 1$ Myr. We also find good agreement with the AD models with  $n_{\rm H} \sim 4\times 10^{5} - 1\times 10^{6}$ cm$^{-3}$, which provide a better fit of the DCN, DNC, and N$_2$H$^+$ abundances. The AD models, however, do a poor job of predicting the CS abundances. One may think that this disagreement is due to our assumption of undepleted sulfur abundance, but this is not the case because assuming sulfur depletion would produce smaller CS abundances and larger discrepancies with observations. Recent fitting of the CS column densities by \citet{Roncero2020} concluded that  we need to assume a sulfur depletion of a factor of 20 and chemical ages $\leq$~1~Myr to account for the observed CS abundance. 

\section{Mock observations: TMC~1-C}
\label{sec:mock}

TMC~1-C has been revealed to be an evolved collapsing core on the verge of forming a star, similar to the well-known L1544 \citep{Caselli2002a, Caselli2012}. Given its relevance in the study of low-mass star formation processes, in this section we carry out a deeper characterization of this core. In Figure~\ref{fig:heatmap} we found that the BES1 model with a central density of $n_{\rm H}=10^{6}$ cm$^{-3}$ fits best the observed molecular abundances in TMC~1-C. AD models with moderate densities $n_{\rm H} = 10^{5}$ cm$^{-3}$ also provide good approximations to the obtained chemical abundances of the species. It is therefore worth investigating how they perform in the radiative transfer of the observed lines. To further test the adequacy of the collapse models presented in the previous section, we post-processed our models to generate the synthetic spectra of a subset of the molecular lines from Table \ref{tab:summarylines}. In order to fully account for the physical structure and the abundance profiles of the different models, we used the radiative transfer code {\scshape Radmc-3d} \citep{RADMC3D}. 

For each density profile we created a spherically symmetric model consisting of a spherical grid divided by spherical shells in an onion-like structure. The spherical shells are endowed with a velocity field following the law $v\sim \alpha\times r^{-0.5}$, where $\alpha$ is chosen to yield line widths similar to those of the observed lines once we take into account the thermal and non-thermal line widths. The thickness of each shell follows a logarithmic law, with thinner layers near the origin and coarser layers towards the edge of the grid. The number of shells was chosen so that the Doppler shift between a layer and its closest neighbors is smaller than the total width of the lines, preventing abnormal ray-tracing (see ``Doppler catching'' in the {\scshape Radmc-3d} documentation). The adopted non-thermal dispersion is $\sigma_{\rm NT}$ = 0.1~km~s$^{-1}$, which fits the line widths of our observations and is in agreement with the results in \citet{Schnee2007}. Finally, the temperature, density, and chemical abundance profiles are interpolated to assign each cell the corresponding value.

\begin{figure*}
	\centering
	\includegraphics[width=\textwidth,keepaspectratio]{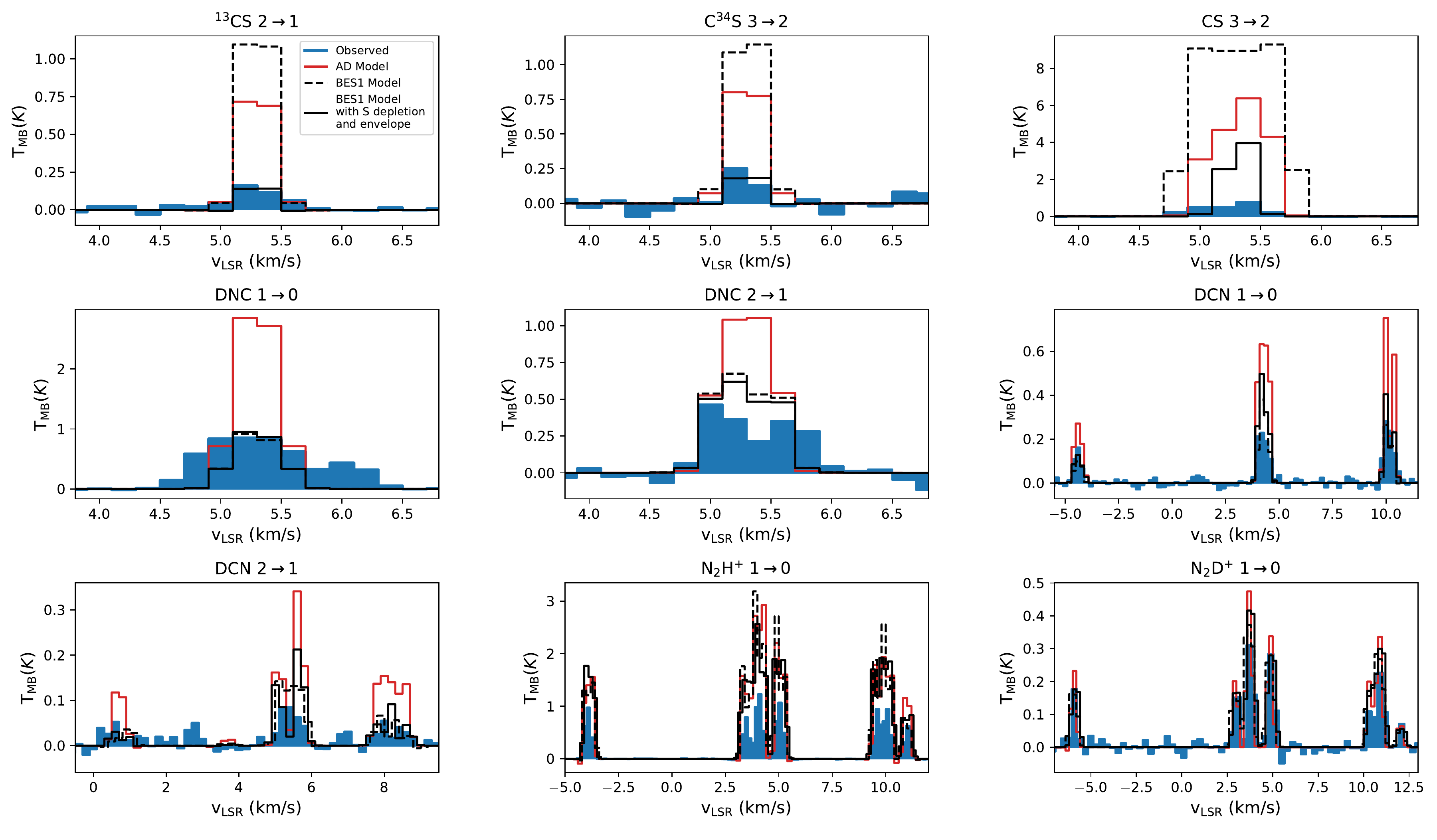}
	\caption{Observed spectra (blue), and the synthetic spectra of different transitions for the AD (red) and BES1 models (black, solid and dashed lines), as results from {\scshape Radmc-3d}. The BES1 model is highlighted as it is the one that, in general, fits the observations better.}
	\label{fig:radmc_models}
\end{figure*}

To estimate the level populations, we used the {\scshape Radmc-3d} Large Velocity Gradient approximation (LVG) implementation \citep{Shetty2011a, Shetty2011b} with the same collision coefficients as in the multi-transition study of Section 5. In Figure~\ref{fig:radmc_models} we show the results of the line radiative transfer of different models following the LVG mode of {\scshape Radmc-3d} after convolving them with the appropriate beam sizes (Table \ref{tab:summarylines}). We first compared the performance of pure BES1 model with central density $n_{\rm H} = 10^{6}$ cm$^{-3}$ and the AD model with $n_{\rm H} = 10^{5}$ cm$^{-3}$. While the BES1 and AD model overestimate the temperatures of the observed CS and isotopolog line temperatures, the AD model performs worse at describing the majority of deuterated lines except N$_{2}$D$^{+}$, where both show good agreement with the observations. It is worth noting that the difference between the synthesized and observed profiles of the DNC $1\rightarrow 0$ line, the observed profile being wider with a secondary component, arises because the hyperfine splitting is not considered in the calculations. We are unable to reproduce these features, due to the absence of collisional coefficient files with hyperfine splitting. In the case of the N$_{2}$H$^{+}$ $1\rightarrow 0$ line, both BES1 and AD models perform poorly. This poor performance suggests uncertainties in the dynamics of the core or in its physical structure. The low observed intensities of the optically thick N$_2$H$^+$ and DCN lines might be due to the existence of a thick absorbing layer that is not accounted for by our profiles. The existence of this thick absorbing layer would also be consistent with the deep self-absorption features detected by \citet{Fuente2019} in the HCN $1\rightarrow 0$ line. To test this idea we added an envelope to the BES1 profile by setting the density, temperature, and chemical abundances constant beyond the point where the density reaches $n_{\rm H_{2}} = 10^{4}$ cm$^{-3}$. The thickness of this envelope was chosen so that the N$_{2}$H$^{+}$ column density would yield the observed opacity of the N$_{2}$H$^{+}$ $1\rightarrow 0$ line. The resulting spectra is shown against the observations in Figure~\ref{fig:radmc_models}. From the results we conclude that the presence of an envelope provides a better agreement to the observations,  but still overestimates the N$_{2}$H$^{+}$ line temperatures by a factor of $\sim 1.5$.

Thus far, the models do a poor job of reproducing the spectra of CS and isotopologs. As noted above, there is a large uncertainty on the value of the initial sulfur abundance. Following the most recent results in \citet{Roncero2020}, we introduced a sulfur depletion of a factor of 20 in the BES1 model with envelope. The resulting line profiles are found to provide the best agreement to the CS $3\rightarrow 2$ line and excellent agreement to  the  J=$3\rightarrow 2$ line of C$^{34}$S and the J=$2\rightarrow 1$ line of $^{13}$CS. Thus, we adopt the BES1 model with a density of $n_{\rm H} = 10^{6}$ cm$^{-3}$ and sulfur depletion as the one that best describes the chemical abundances and spectra of the TMC~1-C starless core.  

Although our model is successful at reproducing the observed spectra, we should use caution before concluding about the collapse timescale in TMC~1-C because there are still some uncertainties that should be considered. On the one hand, our single-dish observations provide beam-averaged ($\sim$4000$-$5000~au) physical and chemical properties, hindering the variations in the density profile and chemical signatures within the core. Higher angular resolution dust continuum emission maps of TMC~1-C show an elongated structure and inhomogeneities within the core that are taken into account in our model \citep{Schnee2005, Schnee2010}. Clumpiness of the densest gas ($n_{\rm H}>10^5$ cm$^{-3}$) within the 30m telescope beam would produce lower brightness temperatures and help reconcile N$_{2}$H$^{+}$ predictions and observations. There is still room for improvement. As suggested by \citet{Sipila2018}, it is necessary to couple the chemistry with the dynamics to reproduce the molecular abundances in dense and small regions formed at the latest stages of the core evolution. Finally, our analysis helped constrain several parameters that are generally poorly known such as $\zeta({\rm H_2})$, the initial OPR of H$_2$, or the initial S abundance. The influence of other initial abundances and quantities, such as the initial C/O ratio, was not taken into account in our study, and could be further constrained with the observation of a larger number of species. 

\section{Discussion}
\label{sec:diss}

The classification of starless cores according to their evolutionary stage has been revealed to be a formidable problem. An accurate measurement of the gas density at the core center is challenging since the abundances of all molecules are depleted and the emerging line intensities are weak due to the low kinetic temperature. Moreover, the emission from the center can be absorbed by the outer lower density layers of the envelope. Chemistry has become an essential tool in determining the physical conditions along these dense cores, and chemical diagnostics such as the N$_2$D$^+$/N$_2$H$^+$ have been successfully used to classify pre-stellar cores. Although extremely useful, chemical models show that all these diagnostics depend on parameters that are usually not well constrained, such as $\zeta({\rm H}_2)$, the initial OPR of H$_2$, and the initial elemental abundances. 

 We used spectroscopic observations to investigate the evolutionary stage of the cores TMC~1-C and TMC~1-CP, located in the nearby region of Taurus. These cores are located in the same molecular cloud at a separation of $\sim 19'$, and are expected to share the same initial conditions. Furthermore, the acquired knowledge of the initial gas ionization degree and chemical composition by \citet{Fuente2019} and \citet{NavarroAlmaida2020} allowed us to carry out a more precise modeling of their chemistry. In addition to species commonly used to probe the evolutionary stage of starless cores such as CS, N$_2$H$^+$, and N$_2$D$^+$, we mapped the deuterated isomers DCN and DNC to further constrain the physical and chemical conditions of the dense gas. The inclusion of these two isomers provided valuable information for the modeling. First, the comparison of the abundances of deuterated neutral and ionic species allowed us to have an estimate of $\zeta({\rm H}_2)$ in the dense gas. Interestingly, the value derived from these species, $\zeta({\rm H}_2)=1.3\times 10^{-17}$ s$^{-1}$, is a factor of 5$-$10 lower than that derived by \citet{Fuente2019} in the translucent cloud. Moreover,  the chemistry of these deuterated isomers is very sensitive to the physical conditions of starless cores. At low temperatures $T_{\rm k} <$ 10 K, the DNC/DCN abundance ratio is a good tracer of the gas density, which itself is an indicator of the starless core evolutionary stage. The abundances of DNC and DCN reach the steady state at earlier times than N$_2$D$^+$. Combining the three species, we are able to distinguish chemical ages ranging from $\sim$0.1 to $\sim$1~Myr.

In an attempt to characterize the contraction of starless cores in TMC~1, we used analytic approximations to the density evolution of starless cores to predict the evolution of the gas chemistry during the collapse. In particular we considered three collapse models: i) the BES1 model, which describes a gravitational collapse with a characteristic time, $t \sim 1$ Myr; ii) the BES4 model, which describes a rapid collapse, $t \sim 0.1$ Myr, induced by the cloud compression; and iii) the AD model, which describes a collapse slowed down by a magnetic support, with $t \sim 10$ Myr. We find that the BES4 model is not compatible with the observations. The abundances measured towards TMC~1-C and TMC~1-CP can only be explained with the BES1 and AD models. Although the BES1 and AD models provide a reasonable fit to the observed chemical abundances, in the case of TMC~1-C  our results are best explained by the BES1 model, which simulates the contraction of a near-equilibrium core with a typical time of collapse of $\sim$1 Myr.

Thus far, only a few starless cores have been studied in detailed using dust continuum emission and molecular spectroscopy. Of these targets, the pre-stellar core L1544, stands out as the prototype of low-mass evolved core on the verge of collapse \citep{Caselli2002a, Caselli2012}; L1544 is a prototypical low-mass pre-stellar core in Taurus. It shows signs of contraction \citep{Caselli2002a, Caselli2012} that would eventually evolve into gravitational collapse, making it an ideal target to study the earliest phases of low-mass star formation. There are many studies that analyze the deuteration in this pre-stellar core \citep[see, e.g.,][]{Caselli2002b, Crapsi2005, Chacon2019, Redaelli2019}. L1544 was also a subject of study in \citet{Crapsi2005}, where they estimate a N$_{2}$H$^{+}$ deuteration fraction of $0.23\pm 0.04$, in accordance with other works. More recently,  \citet{Redaelli2019} estimated the N$_{2}$H$^{+}$ and N$_{2}$D$^{+}$ column densities at the dust peak, yielding  N$_{2}$D$^{+}$/N$_{2}$H$^{+}\approx 0.19^{+0.16}_{-0.19}$. This value is consistent with that measured towards TMC~1-C, suggesting that L1544 and TMC1-C are at a similar evolutionary stage.  \citet{Crapsi2005} carried out a N$_2$H$^+$ and N$_2$D$^+$ survey towards 31 low-mass stars using the 30m telescope. Out of this large sample, only five targets presents N$_2$H$^+$ deuterium fractions similar to L1544 and TMC~1-C, which testifies to the exceptionality of this evolved object that deserves further study. 

The observation of several deuterated species together with N$_2$H$^+$ and CS allowed us to constrain the chemistry of the TMC~1-C. However, there are uncertainties in the physical structure of this core that must be solved for a more accurate description. First of all, the moderate spatial resolution of our observations ($\sim$4000$-$5000~au) hinders the detection of the density and chemical gradients expected within the cores. High spatial resolution molecular observations are mandatory to reveal their complex morphology, density, and chemical structure. At the scales probed by interferometric observations ($<$4000$-$5000~au), the abundances predicted by our pseudo-dynamic model would not be reliable, and a dynamical model is necessary to describe the gas chemical evolution. The comparison of high spatial resolution molecular observations with state-of-the-art 3D chemo-magnetohydrodynamical simulations is necessary to progress in our understanding of the formation and evolution of starless cores.

\section{Summary and conclusions}
Our goal was to investigate the evolution of starless cores based on observations towards the prototypical dense cores TMC~1-C and TMC~1-CP. These two cores are located in the same molecular cloud and share similar initial conditions. Located at a distance of 141.8 pc, these cores provide an excellent opportunity to understand the physical and chemical evolution in this phase.  Our results can be summarized as follows:

\begin{itemize}
\item We mapped the prototypical dense cores TMC~1-C and TMC~1-CP in the J=$2\rightarrow1$ and $3\rightarrow2$ lines of CS (and its isotopologs); the J=$1\rightarrow0$ and $2\rightarrow1$ lines of DCN, DNC, and DN$^{13}$C; and the J= $1\rightarrow0$ line of N$_2$H$^+$ and N$_2$D$^+$. The emission of DNC $1\rightarrow0$ has revealed to be an excellent probe to identify the densest region within these cores. Moreover, we detected DN$^{13}$C $1\rightarrow0$ and $2\rightarrow1$ lines towards TMC~1-C. Based on these observations, we selected five positions for a detailed study. 

\item A multi-transitional study of CS (and isotopologs), DCN, and DNC allowed us to estimate the density and molecular abundances towards the selected positions. The densities derived from CS (and isotopologs) range from $n_{\rm H}$=6$\times$10$^4$~cm$^{-3}$ in TMC~1-CP to $n_{\rm H}$=1$\times$10$^5$ cm$^{-3}$. Based on DNC and DCN observations, we derive $n_{\rm H}$=($4\pm 1$)$\times$10$^4$~cm$^{-3}$ towards all positions. The abundances of deuterated compounds are higher in TMC~1-C. In particular, the N$_2$D$^+$/N$_2$H$^+$ ratio is $\sim$ 0.14 in TMC~1-C and $<0.06$ in TMC~1-CP.  

\item We modeled the chemistry using a gas-grain network that includes deuterium with implemented spin chemistry. As a first step, we explored the parameter space using in a 0D modeling. Our model is able to reproduce the observed abundances, assuming $\zeta(H_2)$=1.3$\times$10$^{-17}$ s$^{-1}$ and a time of $\sim$0.8 Myr for TMC~1-CP and $\sim$2 Myr for TMC~1-C. The initial OPR ratio does not have a strong impact on our calculations since we only consider evolution times higher than 0.1 Myr, for which the OPR evolves to $10^{-3}$. 

\item By using analytic approximations to describe density evolution starless cores as a physical basis for our chemical modeling, we conclude that the TMC~1 filament is not undergoing a rapid collapse. Our results are more consistent with the model BES1 of \citet{Aikawa2005}, which simulates the contraction of a near-equilibrium core with a typical time of collapse of $\sim$1 Myr. TMC~1-CP is in an earlier evolutionary stage due to a shorter time since formation or, alternatively, due to a collapse led by ambipolar diffusion in a magnetically supported core.

\item The 3D radiative transfer of the chemo-dynamical profiles presented in this work supports our conclusions about TMC~1-C. The comparison of the observed spectra with the radiative transfer of the chemo-dynamical profiles from \citet{Aikawa2005} points to a dynamical behavior best described by the BES1 model. This analysis also allowed us to conclude that sulfur-bearing molecular lines are best reproduced with a factor 20 of initial sulfur depletion, in agreement with the most recent results on the subject \citep{Roncero2020}. Finally, the radiative transfer modeling of the N$_{2}$H$^{+}$ $1\rightarrow 0$ line suggests the possible existence of an absorbing envelope in front of this core, which could explain why the observed peak temperatures are lower than expected.   

\end{itemize}
We investigated the chemistry of the starless cores TMC~1-C and TMC~1-CP, which are thought to be in a different evolutionary stage, using CS, N$_2$H$^+$, and the deuterated compounds N$_2$D$^+$, DCN, and DNC as chemical diagnostics. Our data allowed us to reject the possibility that these cores are undergoing a rapid collapse. Moreover, the 3D radiative transfer of pseudo time-dependent collapse models allowed us to discard a collapse timescale longer than $\sim$ 1 Myr for TMC 1-C. However, this analysis does not allow us to definitely differentiate between a nearly equilibrium contraction with a timescale shorter than $\sim$ 1 Myr or a longer collapse slowed down by a magnetic support in TMC 1-CP. A full chemo-magnetohydrodynamical modeling, high angular resolution observations, and an accurate estimate of the magnetic flux could help us figure out a more accurate picture of the collapse history in these regions.

\begin{acknowledgements}
We thank the Spanish MINECO for funding support from AYA2016-75066-C2-1/2-P, PID2019-106235GB-I100. VW acknowledges the CNRS program ``Physique et Chimie du Milieu Interestellaire'' (PCMI), co-funded by the Centre Nationale d'Etudes Spatiales (CNES). S.T.P.M. acknowledge to the European Union's Horizon 2020 research and innovation program for funding support given under grant agreement No. 639459 (PROMISE) and Chalmers Gender Initiative for Excellence (GenIE). I. J. S. has received partial support from the Spanish State Research Agency (AEI; project number PID2019-105552RB-C41).
\end{acknowledgements}

\bibliography{Draft_deut.bib}
\bibliographystyle{aa}

\begin{appendix}
\onecolumn
\section{Line properties.}

In this section we show the obtained integrated areas and main-beam temperatures of the DCN $1\rightarrow 0$, DCN $2\rightarrow 1$, DNC $1\rightarrow 0$, DNC $2\rightarrow 1$, DN$^{13}$C $1\rightarrow 0$, DN$^{13}$C $2\rightarrow 1$, N$_{2}$D$^{+}$ $1\rightarrow 0$, and N$_{2}$H$^{+}$ $1\rightarrow 0$ lines at the different positions listed in Table \ref{tab:targets}. The frequencies and line properties are from the CDMS catalog, except those of the DN$^{13}$C lines, which are from SLAIM catalogue. We also show the results of the HFS-CLASS method applied to the DCN, N$_{2}$H$^{+}$, and N$_{2}$D$^{+}$.
\\
\vspace*{\fill}
\begin{table}[!hb]
		\centering
		\caption{Properties, main beam temperatures, and integrated intensities of the detected spectral lines in TMC~1-C}
			\begin{tabular}{lccccccccc}
				\toprule
				\multirow{2}{*}{\textbf{Species}} &  \multirow{2}{*}{\textbf{Transition}} &  \textbf{Frequency} & \multirow{2}{*}{\textbf{E$_{\rm\mathbf{up}}$} \textbf{(K)}} & \multirow{2}{*}{\textbf{log(A$_{\rm\textbf{ij}}$)}} & \multirow{2}{*}{$\bm{\theta}_{\rm\mathbf{MB}}$ \textbf{(")}} & \multirow{2}{*}{\textbf{g}$_{\rm\mathbf{up}}$} & & \multirow{2}{*}{\textbf{T}$_{\rm\mathbf{MB}}$ \textbf{(K)}} & $\int$\textbf{T}$_{\rm \mathbf{MB}}\ \mathbf{dv}$\\
					
					&  & \textbf{(MHz)} & & & & & & & \textbf{(K km s}$^{\mathbf{-1}}$\textbf{)} \\
				 \midrule\midrule
				
					 \multirow{3}{*}{DCN} & $1_{1}\rightarrow 0_{1}$   & 72413.50 & \multirow{3}{*}{3.5} & -4.88 & \multirow{3}{*}{34} & 3 & & $0.289\pm 0.015$ & $0.164\pm 0.008$ \\
					 & $1_{2}\rightarrow 0_{1}$ & 72414.93 & & -4.88 & & 5 & & $0.249\pm 0.015$ & $0.166\pm 0.008$\\
					 & $1_{0}\rightarrow 0_{1}$ & 72417.03 & & -4.88 & & 1 & & $0.170\pm 0.015$ & $0.073\pm 0.006$\\\midrule
					\multirow{6}{*}{DCN} & $2_{2}\rightarrow 1_{2}$ & 144826.58 & \multirow{6}{*}{10.4} & -4.50 & \multirow{6}{*}{17} & 5 & \rdelim\}{6}{4mm} & \multirow{6}{*}{$0.085\pm 0.019$} & \multirow{6}{*}{$0.056\pm 0.009$}\\
					& $2_{1}\rightarrow 1_{0}$ & 144826.82 & & -4.15 & & 3 & & \\
					& $2_{2}\rightarrow 1_{1}$ & 144828.00 & & -4.02 & & 5 & & \\
					& $2_{3}\rightarrow 1_{2}$ & 144828.11 & & -3.90 & & 7 & & \\
					& $2_{1}\rightarrow 1_{2}$ & 144828.91 & & -5.45 & & 3 & & \\
					& $2_{1}\rightarrow 1_{1}$ & 144830.33 & & -4.28 & & 3 & & & \\
					\midrule
					DNC & $1\rightarrow 0$ & 76305.70 & 3.7 & -4.80 & 34 & 3 & & $0.870\pm 0.010$ & $1.009\pm 0.007$\\
					DNC & $2\rightarrow 1$ & 152609.74 & 11.0 & -3.81 & 17 & 5 & & $0.382\pm 0.048$ & $0.365\pm 0.029$\\
					\midrule
					DN$^{13}$C	&$1\rightarrow 0$ & 73367.7540    &   3.4.    & -4.90  & 34 & 3  &  &   $0.068\pm 0.013$  & $0.098\pm 0.022$  \\		
					DN$^{13}$C	& $2\rightarrow 1$ & 146734.0020 &   10.25 & -3.92  & 27 & 5  &  &   $0.022\pm 0.005$  & $0.063\pm 0.021$  \\
                                          \midrule
					\multirow{15}{*}{N$_{2}$D$^{+}$} & $1_{1,0}\rightarrow 0_{1,1}$ & 77107.46 & \multirow{15}{*}{3.7} & -4.69 & \multirow{15}{*}{34} & 1 & \rdelim\}{6}{4mm} & \multirow{6}{*}{$0.199\pm 0.027$} & \multirow{6}{*}{$0.152\pm 0.014$}\\
					& $1_{1,2}\rightarrow 0_{1,1}$ & 77107.76 & & -5.50 & & 5 & & \\
					& $1_{1,2}\rightarrow 0_{1,2}$ & 77107.76 & & -4.76 & & 5 & \\
					& $1_{1,1}\rightarrow 0_{1,0}$ & 77107.90 & & -4.98 & & 3\\
					& $1_{1,1}\rightarrow 0_{1,1}$ & 77107.90 & & -5.45 & & 3\\
					& $1_{1,1}\rightarrow 0_{1,2}$ & 77107.90 & & -5.61 & & 3\\
					& $1_{2,2}\rightarrow 0_{1,1}$ & 77109.32 & & -5.11 & & 5 & \rdelim\}{2}{4mm} & \multirow{2}{*}{$0.302\pm 0.027$} & \multirow{2}{*}{$0.126\pm 0.009$}\\
					& $1_{2,2}\rightarrow 0_{1,2}$ & 77109.32 & & -5.50 & & 5 & \\
					& $1_{2,3}\rightarrow 0_{1,2}$ & 77109.61 & & -4.69 & & 7 & \rdelim\}{4}{4mm} & \multirow{4}{*}{$0.352\pm 0.026$} & \multirow{4}{*}{$0.151\pm 0.014$}\\
					& $1_{2,1}\rightarrow 0_{1,0}$ & 77109.81 & & -5.19 & & 3 & \\
					& $1_{2,1}\rightarrow 0_{1,1}$ & 77109.81 & & -4.87 & & 3\\
					& $1_{2,1}\rightarrow 0_{1,2}$ & 77109.81 & & -6.09 & & 3\\
					& $1_{0,1}\rightarrow 0_{1,1}$ & 77112.11 & & -5.32 & & 3 & \rdelim\}{3}{4mm} & \multirow{3}{*}{$0.181\pm 0.026$} & \multirow{3}{*}{$0.072\pm 0.014$}\\
					& $1_{0,1}\rightarrow 0_{1,2}$ & 77112.11 & & -4.92 & & 3\\
					& $1_{0,1}\rightarrow 0_{1,0}$ & 77112.11 & & -5.44 & & 3\\
					\midrule
					\multirow{15}{*}{N$_{2}$H$^{+}$} & $1_{1,0}\rightarrow 0_{1,1}$ & 93171.62 & \multirow{15}{*}{4.5} & -4.44 & \multirow{15}{*}{26} & 1 &  & $0.640\pm 0.025$ & $0.219\pm 0.035$\\
					& $1_{1,2}\rightarrow 0_{1,1}$ & 93171.91 & & -5.25 & & 5 & \rdelim\}{5}{4mm} & \multirow{5}{*}{$0.734\pm 0.025$} & \multirow{5}{*}{$0.582\pm 0.035$}\\
					& $1_{1,2}\rightarrow 0_{1,2}$ & 93171.91 & & -4.51 & & 5\\
					& $1_{1,1}\rightarrow 0_{1,0}$ & 93172.05 & & -4.73 & & 3\\
					& $1_{1,1}\rightarrow 0_{1,1}$ & 93172.05 & & -5.36 & & 3\\
					& $1_{1,1}\rightarrow 0_{1,2}$ & 93172.05 & & -4.87 & & 3\\
					& $1_{2,2}\rightarrow 0_{1,1}$ & 93173.47 & & -4.51 & & 5 & \rdelim\}{2}{4mm} & \multirow{2}{*}{$0.759\pm 0.025$} & \multirow{2}{*}{$0.293\pm 0.035$}\\
					& $1_{2,2}\rightarrow 0_{1,2}$ & 93173.47 & & -5.25 & & 5\\
					& $1_{2,3}\rightarrow 0_{1,2}$ & 93173.77 & & -4.44 & & 7 & \rdelim\}{4}{4mm} & \multirow{4}{*}{$0.610\pm 0.025$} & \multirow{4}{*}{$0.658\pm 0.035$}\\
					& $1_{2,1}\rightarrow 0_{1,0}$ & 93173.96 & & -4.95 & & 3 & \\
					& $1_{2,1}\rightarrow 0_{1,1}$ & 93173.96 & & -4.63 & & 3\\
					& $1_{2,1}\rightarrow 0_{1,2}$ & 93173.96 & & -5.84 & & 3\\
					& $1_{0,1}\rightarrow 0_{1,0}$ & 93176.26 & & -5.19 & & 3 & \rdelim\}{3}{4mm} & \multirow{3}{*}{$0.811\pm 0.025$} & \multirow{3}{*}{$0.282\pm 0.035$}\\
					& $1_{0,1}\rightarrow 0_{1,1}$ & 93176.26 & & -5.07 & & 3\\
					& $1_{0,1}\rightarrow 0_{1,2}$ & 93176.26 & & -4.67 & & 3\\
					\midrule
    			\bottomrule
			\end{tabular}
			\flushleft
			{\small
			}
			\label{tab:lines1}
\end{table}
\vspace*{\fill}
\newpage
\vspace*{\fill}
\begin{table}[!hb]
		\caption{Properties, main beam temperatures, and integrated intensities of the detected spectral lines in TMC~1-C (P)}
			\begin{tabular}{lccccccccc}
				\toprule
				\multirow{2}{*}{\textbf{Species}} &  \multirow{2}{*}{\textbf{Transition}} &  \textbf{Frequency} & \multirow{2}{*}{\textbf{E$_{\rm\mathbf{up}}$} \textbf{(K)}} & \multirow{2}{*}{\textbf{log(A$_{\rm\textbf{ij}}$)}} & \multirow{2}{*}{$\bm{\theta}_{\rm\mathbf{MB}}$ \textbf{(")}} & \multirow{2}{*}{\textbf{g}$_{\rm\mathbf{up}}$} & & \multirow{2}{*}{\textbf{T}$_{\rm\mathbf{MB}}$ \textbf{(K)}} & $\int$\textbf{T}$_{\rm \mathbf{MB}}\ \mathbf{dv}$\\
					
					&  & \textbf{(MHz)} & & & & & & & \textbf{(K km s}$^{\mathbf{-1}}$\textbf{)} \\
				 \midrule\midrule
				
					 \multirow{3}{*}{DCN} & $1_{1}\rightarrow 0_{1}$   & 72413.50 & \multirow{3}{*}{3.5} & -4.88 & \multirow{3}{*}{34} & 3 & & $0.408\pm 0.012$ & $0.185\pm 0.007$ \\
					 & $1_{2}\rightarrow 0_{1}$ & 72414.93 & & -4.88 & & 5 & & $0.484\pm 0.012$ & $0.243\pm 0.007$\\
					 & $1_{0}\rightarrow 0_{1}$ & 72417.03 & & -4.88 & & 1 & & $0.271\pm 0.012$ & $0.091\pm 0.006$\\\midrule
					\multirow{6}{*}{DCN} & $2_{2}\rightarrow 1_{2}$ & 144826.58 & \multirow{6}{*}{10.4} & -4.50 & \multirow{6}{*}{17} & 5 & \rdelim\}{2}{4mm} & \multirow{2}{*}{$0.091\pm 0.026$} & \multirow{2}{*}{$0.045\pm 0.015$}\\
					& $2_{1}\rightarrow 1_{0}$ & 144826.82 & & -4.15 & & 3 & & \\
					& $2_{2}\rightarrow 1_{1}$ & 144828.00 & & -4.02 & & 5 & \rdelim\}{3}{4mm} & \multirow{3}{*}{$0.130\pm 0.026$} & \multirow{3}{*}{$0.079\pm 0.010$}\\
					& $2_{3}\rightarrow 1_{2}$ & 144828.11 & & -3.90 & & 7\\
					& $2_{1}\rightarrow 1_{2}$ & 144828.91 & & -5.45 & & 3\\
					& $2_{1}\rightarrow 1_{1}$ & 144830.33 & & -4.28 & & 3 & & $0.064\pm 0.026$ & $0.058\pm 0.013$\\
					\midrule
					DCN & $3\rightarrow 2$ & 217237.00 & 20.9 & -3.34 & 11 & 21 & & $0.071\pm 0.022$ & $0.027\pm 0.009$\\
					\midrule
					DNC & $1\rightarrow 0$ & 76305.70 & 3.7 & -4.80 & 34 & 3 & & $1.387\pm 0.120$ & $1.299\pm 0.009$\\
					DNC & $2\rightarrow 1$ & 152609.74 & 11.0 & -3.81 & 17 & 5 & & $0.726\pm 0.089$ & $0.557\pm 0.018$\\
					\midrule
					\multirow{15}{*}{N$_{2}$D$^{+}$} & $1_{1,0}\rightarrow 0_{1,1}$ & 77107.46 & \multirow{15}{*}{3.7} & -4.69 & \multirow{15}{*}{34} & 1 & \rdelim\}{6}{4mm} & \multirow{6}{*}{$0.249\pm 0.033$} & \multirow{6}{*}{$0.098\pm 0.011$}\\
					& $1_{1,2}\rightarrow 0_{1,1}$ & 77107.76 & & -5.50 & & 5 \\
					& $1_{1,2}\rightarrow 0_{1,2}$ & 77107.76 & & -4.76 & & 5 & \\
					& $1_{1,1}\rightarrow 0_{1,0}$ & 77107.90 & & -4.98 & & 3\\
					& $1_{1,1}\rightarrow 0_{1,1}$ & 77107.90 & & -5.45 & & 3\\
					& $1_{1,1}\rightarrow 0_{1,2}$ & 77107.90 & & -5.61 & & 3\\
					& $1_{2,2}\rightarrow 0_{1,1}$ & 77109.32 & & -5.11 & & 5 & \rdelim\}{2}{4mm} & \multirow{2}{*}{$0.271\pm 0.033$} & \multirow{2}{*}{$0.096\pm 0.012$}\\
					& $1_{2,2}\rightarrow 0_{1,2}$ & 77109.32 & & -5.50 & & 5 & \\
					& $1_{2,3}\rightarrow 0_{1,2}$ & 77109.61 & & -4.69 & & 7 & \rdelim\}{4}{4mm} & \multirow{4}{*}{$0.361\pm 0.033$} & \multirow{4}{*}{$0.136\pm 0.012$}\\
					& $1_{2,1}\rightarrow 0_{1,0}$ & 77109.81 & & -5.19 & & 3\\
					& $1_{2,1}\rightarrow 0_{1,1}$ & 77109.81 & & -4.87 & & 3\\
					& $1_{2,1}\rightarrow 0_{1,2}$ & 77109.81 & & -6.09 & & 3\\
					& $1_{0,1}\rightarrow 0_{1,1}$ & 77112.11 & & -5.32 & & 3 & \rdelim\}{3}{4mm} & \multirow{3}{*}{$0.152\pm 0.033$} & \multirow{3}{*}{$0.048\pm 0.011$}\\
					& $1_{0,1}\rightarrow 0_{1,2}$ & 77112.11 & & -4.92 & & 3\\
					& $1_{0,1}\rightarrow 0_{1,0}$ & 77112.11 & & -5.44 & & 3\\
					\midrule
					\multirow{15}{*}{N$_{2}$H$^{+}$} & $1_{1,0}\rightarrow 0_{1,1}$ & 93171.62 & \multirow{15}{*}{4.5} & -4.44 & \multirow{15}{*}{26} & 1 & & $0.670\pm 0.023$ & $0.198\pm 0.009$\\
					& $1_{1,2}\rightarrow 0_{1,1}$ & 93171.91 & & -5.25 & & 5 & \rdelim\}{5}{4mm} & \multirow{5}{*}{$0.803\pm 0.023$} & \multirow{5}{*}{$0.623\pm 0.052$}\\
					& $1_{1,2}\rightarrow 0_{1,2}$ & 93171.91 & & -4.51 & & 5\\
					& $1_{1,1}\rightarrow 0_{1,0}$ & 93172.05 & & -4.73 & & 3\\
					& $1_{1,1}\rightarrow 0_{1,1}$ & 93172.05 & & -5.36 & & 3\\
					& $1_{1,1}\rightarrow 0_{1,2}$ & 93172.05 & & -4.87 & & 3\\
					& $1_{2,2}\rightarrow 0_{1,1}$ & 93173.47 & & -4.51 & & 5 & \rdelim\}{2}{4mm} & \multirow{2}{*}{$1.050\pm 0.023$} & \multirow{2}{*}{$0.323\pm 0.010$}\\
					& $1_{2,2}\rightarrow 0_{1,2}$ & 93173.47 & & -5.25 & & 5\\
					& $1_{2,3}\rightarrow 0_{1,2}$ & 93173.77 & & -4.44 & & 7 & \rdelim\}{4}{4mm} & \multirow{4}{*}{$0.711\pm 0.023$} & \multirow{4}{*}{$0.708\pm 0.015$}\\
					& $1_{2,1}\rightarrow 0_{1,0}$ & 93173.96 & & -4.95 & & 3\\
					& $1_{2,1}\rightarrow 0_{1,1}$ & 93173.96 & & -4.63 & & 3\\
					& $1_{2,1}\rightarrow 0_{1,2}$ & 93173.96 & & -5.84 & & 3\\
					& $1_{0,1}\rightarrow 0_{1,0}$ & 93176.26 & & -5.19 & & 3 & \rdelim\}{3}{4mm} & \multirow{3}{*}{$0.946\pm 0.023$} & \multirow{3}{*}{$0.320\pm 0.010$}\\
					& $1_{0,1}\rightarrow 0_{1,1}$ & 93176.26 & & -5.07 & & 3\\
					& $1_{0,1}\rightarrow 0_{1,2}$ & 93176.26 & & -4.67 & & 3\\
					\midrule
    			\bottomrule
			\end{tabular}
			\label{tab:lines2}
\end{table}
\vspace*{\fill}
\newpage
\vspace*{\fill}
\begin{table}[!hb]
		\caption{Properties, main beam temperatures, and integrated intensities of the detected spectral lines in TMC~1-CP}
			\resizebox{\textwidth}{!}{
			\begin{tabular}{lccccccccc}
				\toprule
				\multirow{2}{*}{\textbf{Species}} &  \multirow{2}{*}{\textbf{Transition}} &  \textbf{Frequency} & \multirow{2}{*}{\textbf{E$_{\rm\mathbf{up}}$} \textbf{(K)}} & \multirow{2}{*}{\textbf{log(A$_{\rm\textbf{ij}}$)}} & \multirow{2}{*}{$\bm{\theta}_{\rm\mathbf{MB}}$ \textbf{(")}} & \multirow{2}{*}{\textbf{g}$_{\rm\mathbf{up}}$} & & \multirow{2}{*}{\textbf{T}$_{\rm\mathbf{MB}}$ \textbf{(K)}} & $\int$\textbf{T}$_{\rm \mathbf{MB}}\ \mathbf{dv}$\\
					
					&  & \textbf{(MHz)} & & & & & & & \textbf{(K km s}$^{\mathbf{-1}}$\textbf{)} \\
				 \midrule\midrule
				
					 \multirow{3}{*}{DCN} & $1_{1}\rightarrow 0_{1}$   & 72413.50 & \multirow{3}{*}{3.5} & -4.88 & \multirow{3}{*}{34} & 3 & & $0.287\pm 0.020$ & $0.192\pm 0.012$ \\
					 & $1_{2}\rightarrow 0_{1}$ & 72414.93 & & -4.88 & & 5 & & $0.461\pm 0.020$ & $0.320\pm 0.011$\\
					 & $1_{0}\rightarrow 0_{1}$ & 72417.03 & & -4.88 & & 1 & & $0.145\pm 0.020$ & $0.074\pm 0.010$\\\midrule
					\multirow{6}{*}{DCN} & $2_{2}\rightarrow 1_{2}$ & 144826.58 & \multirow{6}{*}{10.4} & -4.50 & \multirow{6}{*}{17} & 5 & \rdelim\}{6}{4mm} & \multirow{6}{*}{$0.177\pm 0.014$} & \multirow{6}{*}{$0.093\pm 0.007$}\\
					& $2_{1}\rightarrow 1_{0}$ & 144826.82 & & -4.15 & & 3 & & \\
					& $2_{2}\rightarrow 1_{1}$ & 144828.00 & & -4.02 & & 5 & & \\
					& $2_{3}\rightarrow 1_{2}$ & 144828.11 & & -3.90 & & 7 & & \\
					& $2_{1}\rightarrow 1_{2}$ & 144828.91 & & -5.45 & & 3 & & \\
					& $2_{1}\rightarrow 1_{1}$ & 144830.33 & & -4.28 & & 3 & & & \\
					\midrule
					DNC & $1\rightarrow 0$ & 76305.70 & 3.7 & -4.80 & 34 & 3 & & $0.964\pm 0.012$ & $1.051\pm 0.008$\\
					DNC & $2\rightarrow 1$ & 152609.74 & 11.0 & -3.81 & 17 & 5 & & $0.722\pm 0.058$ & $0.535\pm 0.031$\\
					\midrule
					\multirow{15}{*}{N$_{2}$D$^{+}$} & $1_{1,0}\rightarrow 0_{1,1}$ & 77107.46 & \multirow{15}{*}{3.7} & -4.69 & \multirow{15}{*}{34} & 1 & \rdelim\}{15}{4mm} & \multirow{15}{*}{{\rm rms:} $1.66\times 10^{-2}$ K} & 
					\\
					& $1_{1,2}\rightarrow 0_{1,1}$ & 77107.76 & & -5.50 & & 5 & & & \\
					& $1_{1,2}\rightarrow 0_{1,2}$ & 77107.76 & & -4.76 & & 5 & \\
					& $1_{1,1}\rightarrow 0_{1,0}$ & 77107.90 & & -4.98 & & 3\\
					& $1_{1,1}\rightarrow 0_{1,1}$ & 77107.90 & & -5.45 & & 3\\
					& $1_{1,1}\rightarrow 0_{1,2}$ & 77107.90 & & -5.61 & & 3\\
					& $1_{2,2}\rightarrow 0_{1,1}$ & 77109.32 & & -5.11 & & 5 & 
					\\
					& $1_{2,2}\rightarrow 0_{1,2}$ & 77109.32 & & -5.50 & & 5 & \\
					& $1_{2,3}\rightarrow 0_{1,2}$ & 77109.61 & & -4.69 & & 7\\
					& $1_{2,1}\rightarrow 0_{1,0}$ & 77109.81 & & -5.19 & & 3\\
					& $1_{2,1}\rightarrow 0_{1,1}$ & 77109.81 & & -4.87 & & 3\\
					& $1_{2,1}\rightarrow 0_{1,2}$ & 77109.81 & & -6.09 & & 3\\
					& $1_{0,1}\rightarrow 0_{1,1}$ & 77112.11 & & -5.32 & & 3 & 
					\\
					& $1_{0,1}\rightarrow 0_{1,2}$ & 77112.11 & & -4.92 & & 3 & & 
					\\ 
					& $1_{0,1}\rightarrow 0_{1,0}$ & 77112.11 & & -5.44 & & 3\\
					\midrule
					\multirow{15}{*}{N$_{2}$H$^{+}$} & $1_{1,0}\rightarrow 0_{1,1}$ & 93171.62 & \multirow{15}{*}{4.5} & -4.44 & \multirow{15}{*}{26} & 1 & & $0.427\pm 0.007$ & $0.170\pm 0.053$\\
					& $1_{1,2}\rightarrow 0_{1,1}$ & 93171.91 & & -5.25 & & 5 & \rdelim\}{5}{4mm} & \multirow{5}{*}{$0.955\pm 0.007$} & \multirow{5}{*}{$0.835\pm 0.043$}\\
					& $1_{1,2}\rightarrow 0_{1,2}$ & 93171.91 & & -4.51 & & 5\\
					& $1_{1,1}\rightarrow 0_{1,0}$ & 93172.05 & & -4.73 & & 3\\
					& $1_{1,1}\rightarrow 0_{1,1}$ & 93172.05 & & -5.36 & & 3\\
					& $1_{1,1}\rightarrow 0_{1,2}$ & 93172.05 & & -4.87 & & 3\\
					& $1_{2,2}\rightarrow 0_{1,1}$ & 93173.47 & & -4.51 & & 5 & \rdelim\}{2}{4mm} & \multirow{2}{*}{$1.046\pm 0.007$} & \multirow{2}{*}{$0.509\pm 0.043$}\\
					& $1_{2,2}\rightarrow 0_{1,2}$ & 93173.47 & & -5.25 & & 5\\
					& $1_{2,3}\rightarrow 0_{1,2}$ & 93173.77 & & -4.44 & & 7 & \rdelim\}{4}{4mm} & \multirow{4}{*}{$1.191\pm 0.007$} & \multirow{4}{*}{$1.071\pm 0.043$}\\
					& $1_{2,1}\rightarrow 0_{1,0}$ & 93173.96 & & -4.95 & & 3\\
					& $1_{2,1}\rightarrow 0_{1,1}$ & 93173.96 & & -4.63 & & 3\\
					& $1_{2,1}\rightarrow 0_{1,2}$ & 93173.96 & & -5.84 & & 3\\
					& $1_{0,1}\rightarrow 0_{1,0}$ & 93176.26 & & -5.19 & & 3 & \rdelim\}{3}{4mm} & \multirow{3}{*}{$0.808\pm 0.007$} & \multirow{3}{*}{$0.423\pm 0.043$}\\
					& $1_{0,1}\rightarrow 0_{1,1}$ & 93176.26 & & -5.07 & & 3\\
					& $1_{0,1}\rightarrow 0_{1,2}$ & 93176.26 & & -4.67 & & 3\\
					\midrule
    			\bottomrule
			\end{tabular}
			}
			\label{tab:lines3}
\end{table}
\vspace*{\fill}
\newpage
\vspace*{\fill}
\begin{table}[!hb]
		\caption{Properties, main beam temperatures, and integrated intensities of the detected spectral lines in TMC~1-CP (P1)}
			\resizebox{\textwidth}{!}{
			\begin{tabular}{lccccccccc}
				\toprule
				\multirow{2}{*}{\textbf{Species}} &  \multirow{2}{*}{\textbf{Transition}} &  \textbf{Frequency} & \multirow{2}{*}{\textbf{E$_{\rm\mathbf{up}}$} \textbf{(K)}} & \multirow{2}{*}{\textbf{log(A$_{\rm\textbf{ij}}$)}} & \multirow{2}{*}{$\bm{\theta}_{\rm\mathbf{MB}}$ \textbf{(")}} & \multirow{2}{*}{\textbf{g}$_{\rm\mathbf{up}}$} & & \multirow{2}{*}{\textbf{T}$_{\rm\mathbf{MB}}$ \textbf{(K)}} & $\int$\textbf{T}$_{\rm \mathbf{MB}}\ \mathbf{dv}$\\
					
					&  & \textbf{(MHz)} & & & & & & & \textbf{(K km s}$^{\mathbf{-1}}$\textbf{)} \\
				 \midrule\midrule
				
					 \multirow{3}{*}{DCN} & $1_{1}\rightarrow 0_{1}$   & 72413.50 & \multirow{3}{*}{3.5} & -4.88 & \multirow{3}{*}{34} & 3 & & $0.358\pm 0.008$ & $0.190\pm 0.013$ \\
					 & $1_{2}\rightarrow 0_{1}$ & 72414.93 & & -4.88 & & 5 & & $0.530\pm 0.008$ & $0.330\pm 0.011$\\
					 & $1_{0}\rightarrow 0_{1}$ & 72417.03 & & -4.88 & & 1 & & $0.222\pm 0.008$ & $0.083\pm 0.014$\\\midrule
					\multirow{6}{*}{DCN} & $2_{2}\rightarrow 1_{2}$ & 144826.58 & \multirow{6}{*}{10.4} & -4.50 & \multirow{6}{*}{17} & 5 & \rdelim\}{6}{4mm} & \multirow{6}{*}{$0.282\pm 0.016$} & \multirow{6}{*}{$0.110\pm 0.015$}\\
					& $2_{1}\rightarrow 1_{0}$ & 144826.82 & & -4.15 & & 3 & & \\
					& $2_{2}\rightarrow 1_{1}$ & 144828.00 & & -4.02 & & 5 & & \\
					& $2_{3}\rightarrow 1_{2}$ & 144828.11 & & -3.90 & & 7 & & \\
					& $2_{1}\rightarrow 1_{2}$ & 144828.91 & & -5.45 & & 3 & & \\
					& $2_{1}\rightarrow 1_{1}$ & 144830.33 & & -4.28 & & 3 & & & \\
					\midrule
					DNC & $1\rightarrow 0$ & 76305.70 & 3.7 & -4.80 & 34 & 3 & & $1.227\pm 0.080$ & $1.215\pm 0.013$\\
					DNC & $2\rightarrow 1$ & 152609.74 & 11.0 & -3.81 & 17 & 5 & & $0.732\pm 0.059$ & $0.477\pm 0.027$\\
					\midrule
					\multirow{15}{*}{N$_{2}$D$^{+}$} & $1_{1,0}\rightarrow 0_{1,1}$ & 77107.46 & \multirow{15}{*}{3.7} & -4.69 & \multirow{15}{*}{34} & 1 & \rdelim\}{15}{4mm} & \multirow{15}{*}{{\rm rms:} $2.00\times 10^{-2}$ K} & 
					\\
					& $1_{1,2}\rightarrow 0_{1,1}$ & 77107.76 & & -5.50 & & 5 & & & \\
					& $1_{1,2}\rightarrow 0_{1,2}$ & 77107.76 & & -4.76 & & 5 & \\
					& $1_{1,1}\rightarrow 0_{1,0}$ & 77107.90 & & -4.98 & & 3\\
					& $1_{1,1}\rightarrow 0_{1,1}$ & 77107.90 & & -5.45 & & 3\\
					& $1_{1,1}\rightarrow 0_{1,2}$ & 77107.90 & & -5.61 & & 3\\
					& $1_{2,2}\rightarrow 0_{1,1}$ & 77109.32 & & -5.11 & & 5 & 
					\\
					& $1_{2,2}\rightarrow 0_{1,2}$ & 77109.32 & & -5.50 & & 5 & \\
					& $1_{2,3}\rightarrow 0_{1,2}$ & 77109.61 & & -4.69 & & 7 & 
					\\
					& $1_{2,1}\rightarrow 0_{1,0}$ & 77109.81 & & -5.19 & & 3\\
					& $1_{2,1}\rightarrow 0_{1,1}$ & 77109.81 & & -4.87 & & 3\\
					& $1_{2,1}\rightarrow 0_{1,2}$ & 77109.81 & & -6.09 & & 3\\
					& $1_{0,1}\rightarrow 0_{1,1}$ & 77112.11 & & -5.32 & & 3 & 
					\\
					& $1_{0,1}\rightarrow 0_{1,2}$ & 77112.11 & & -4.92 & & 3 & & \\ 
					& $1_{0,1}\rightarrow 0_{1,0}$ & 77112.11 & & -5.44 & & 3\\
					\midrule
					\multirow{15}{*}{N$_{2}$H$^{+}$} & $1_{1,0}\rightarrow 0_{1,1}$ & 93171.62 & \multirow{15}{*}{4.5} & -4.44 & \multirow{15}{*}{26} & 1 & & $0.379\pm 0.022$ & $0.167\pm 0.008$\\
					& $1_{1,2}\rightarrow 0_{1,1}$ & 93171.91 & & -5.25 & & 5 & \rdelim\}{5}{4mm} & \multirow{5}{*}{$1.061\pm 0.022$} & \multirow{5}{*}{$0.904\pm 0.013$}\\
					& $1_{1,2}\rightarrow 0_{1,2}$ & 93171.91 & & -4.51 & & 5\\
					& $1_{1,1}\rightarrow 0_{1,0}$ & 93172.05 & & -4.73 & & 3\\
					& $1_{1,1}\rightarrow 0_{1,1}$ & 93172.05 & & -5.36 & & 3\\
					& $1_{1,1}\rightarrow 0_{1,2}$ & 93172.05 & & -4.87 & & 3\\
					& $1_{2,2}\rightarrow 0_{1,1}$ & 93173.47 & & -4.51 & & 5 & \rdelim\}{2}{4mm} & \multirow{2}{*}{$1.208\pm 0.022$} & \multirow{2}{*}{$0.542\pm 0.009$}\\
					& $1_{2,2}\rightarrow 0_{1,2}$ & 93173.47 & & -5.25 & & 5\\
					& $1_{2,3}\rightarrow 0_{1,2}$ & 93173.77 & & -4.44 & & 7 & \rdelim\}{4}{4mm} & \multirow{4}{*}{$1.263\pm 0.022$} & \multirow{4}{*}{$1.102\pm 0.015$}\\
					& $1_{2,1}\rightarrow 0_{1,0}$ & 93173.96 & & -4.95 & & 3\\
					& $1_{2,1}\rightarrow 0_{1,1}$ & 93173.96 & & -4.63 & & 3\\
					& $1_{2,1}\rightarrow 0_{1,2}$ & 93173.96 & & -5.84 & & 3\\
					& $1_{0,1}\rightarrow 0_{1,0}$ & 93176.26 & & -5.19 & & 3 & \rdelim\}{3}{4mm} & \multirow{3}{*}{$0.889\pm 0.022$} & \multirow{3}{*}{$0.430\pm 0.010$}\\
					& $1_{0,1}\rightarrow 0_{1,1}$ & 93176.26 & & -5.07 & & 3\\
					& $1_{0,1}\rightarrow 0_{1,2}$ & 93176.26 & & -4.67 & & 3\\
					\midrule
    			\bottomrule
			\end{tabular}
			}
			\label{tab:lines4}
\end{table}
\vspace*{\fill}
\newpage
\vspace*{\fill}
\begin{table}[!hb]
		\centering
		\caption{Properties, main beam temperatures, and integrated intensities of the detected spectral lines in TMC~1-CP (P2)}
			\resizebox{\textwidth}{!}{
			\begin{tabular}{lccccccccc}
				\toprule
				\multirow{2}{*}{\textbf{Species}} &  \multirow{2}{*}{\textbf{Transition}} &  \textbf{Frequency} & \multirow{2}{*}{\textbf{E$_{\rm\mathbf{up}}$} \textbf{(K)}} & \multirow{2}{*}{\textbf{log(A$_{\rm\textbf{ij}}$)}} & \multirow{2}{*}{$\bm{\theta}_{\rm\mathbf{MB}}$ \textbf{(")}} & \multirow{2}{*}{\textbf{g}$_{\rm\mathbf{up}}$} & & \multirow{2}{*}{\textbf{T}$_{\rm\mathbf{MB}}$ \textbf{(K)}} & $\int$\textbf{T}$_{\rm \mathbf{MB}}\ \mathbf{dv}$\\
					
					&  & \textbf{(MHz)} & & & & & & & \textbf{(K km s}$^{\mathbf{-1}}$\textbf{)} \\
				 \midrule\midrule
				
					 \multirow{3}{*}{DCN} & $1_{1}\rightarrow 0_{1}$   & 72413.50 & \multirow{3}{*}{3.5} & -4.88 & \multirow{3}{*}{34} & 3 & & $0.363\pm 0.033$ & $0.220\pm 0.015$ \\
					 & $1_{2}\rightarrow 0_{1}$ & 72414.93 & & -4.88 & & 5 & & $0.542\pm 0.033$ & $0.357\pm 0.018$\\
					 & $1_{0}\rightarrow 0_{1}$ & 72417.03 & & -4.88 & & 1 & & $0.219\pm 0.033$ & $0.090\pm 0.014$\\\midrule
					\multirow{6}{*}{DCN} & $2_{2}\rightarrow 1_{2}$ & 144826.58 & \multirow{6}{*}{10.4} & -4.50 & \multirow{6}{*}{17} & 5 & \rdelim\}{6}{4mm} & \multirow{6}{*}{$0.227\pm 0.021$} & \multirow{6}{*}{$0.182\pm 0.022$}\\
					& $2_{1}\rightarrow 1_{0}$ & 144826.82 & & -4.15 & & 3 & & \\
					& $2_{2}\rightarrow 1_{1}$ & 144828.00 & & -4.02 & & 5 & & \\
					& $2_{3}\rightarrow 1_{2}$ & 144828.11 & & -3.90 & & 7 & & \\
					& $2_{1}\rightarrow 1_{2}$ & 144828.91 & & -5.45 & & 3 & & \\
					& $2_{1}\rightarrow 1_{1}$ & 144830.33 & & -4.28 & & 3 & & & \\
					\midrule
					DNC & $1\rightarrow 0$ & 76305.70 & 3.7 & -4.80 & 34 & 3 & & $1.166\pm 0.016$ & $1.179\pm 0.011$\\
					DNC & $2\rightarrow 1$ & 152609.74 & 11.0 & -3.81 & 17 & 5 & & $0.778\pm 0.048$ & $0.583\pm 0.032$\\
					\midrule
					\multirow{15}{*}{N$_{2}$D$^{+}$} & $1_{1,0}\rightarrow 0_{1,1}$ & 77107.46 & \multirow{15}{*}{3.7} & -4.69 & \multirow{15}{*}{34} & 1 & \rdelim\}{15}{4mm} & \multirow{15}{*}{{\rm rms:} $2.20\times 10^{-2}$ K} & 
					\\
					& $1_{1,2}\rightarrow 0_{1,1}$ & 77107.76 & & -5.50 & & 5 & & & \\
					& $1_{1,2}\rightarrow 0_{1,2}$ & 77107.76 & & -4.76 & & 5 & \\
					& $1_{1,1}\rightarrow 0_{1,0}$ & 77107.90 & & -4.98 & & 3\\
					& $1_{1,1}\rightarrow 0_{1,1}$ & 77107.90 & & -5.45 & & 3\\
					& $1_{1,1}\rightarrow 0_{1,2}$ & 77107.90 & & -5.61 & & 3\\
					& $1_{2,2}\rightarrow 0_{1,1}$ & 77109.32 & & -5.11 & & 5 & 
					\\
					& $1_{2,2}\rightarrow 0_{1,2}$ & 77109.32 & & -5.50 & & 5  \\
					& $1_{2,3}\rightarrow 0_{1,2}$ & 77109.61 & & -4.69 & & 7 & 
					\\
					& $1_{2,1}\rightarrow 0_{1,0}$ & 77109.81 & & -5.19 & & 3\\
					& $1_{2,1}\rightarrow 0_{1,1}$ & 77109.81 & & -4.87 & & 3\\
					& $1_{2,1}\rightarrow 0_{1,2}$ & 77109.81 & & -6.09 & & 3\\
					& $1_{0,1}\rightarrow 0_{1,1}$ & 77112.11 & & -5.32 & & 3 & 
					\\
					& $1_{0,1}\rightarrow 0_{1,2}$ & 77112.11 & & -4.92 & & 3 & & \\ 
					& $1_{0,1}\rightarrow 0_{1,0}$ & 77112.11 & & -5.44 & & 3\\
					\midrule
					\multirow{15}{*}{N$_{2}$H$^{+}$} & $1_{1,0}\rightarrow 0_{1,1}$ & 93171.62 & \multirow{15}{*}{4.5} & -4.44 & \multirow{15}{*}{26} & 1 & & $0.347\pm 0.019$ & $0.133\pm 0.052$\\
					& $1_{1,2}\rightarrow 0_{1,1}$ & 93171.91 & & -5.25 & & 5 & \rdelim\}{5}{4mm} & \multirow{5}{*}{$1.005\pm 0.019$} & \multirow{5}{*}{$0.885\pm 0.051$}\\
					& $1_{1,2}\rightarrow 0_{1,2}$ & 93171.91 & & -4.51 & & 5\\
					& $1_{1,1}\rightarrow 0_{1,0}$ & 93172.05 & & -4.73 & & 3\\
					& $1_{1,1}\rightarrow 0_{1,1}$ & 93172.05 & & -5.36 & & 3\\
					& $1_{1,1}\rightarrow 0_{1,2}$ & 93172.05 & & -4.87 & & 3\\
					& $1_{2,2}\rightarrow 0_{1,1}$ & 93173.47 & & -4.51 & & 5 & \rdelim\}{2}{4mm} & \multirow{2}{*}{$1.151\pm 0.022$} & \multirow{2}{*}{$0.502\pm 0.051$}\\
					& $1_{2,2}\rightarrow 0_{1,2}$ & 93173.47 & & -5.25 & & 5\\
					& $1_{2,3}\rightarrow 0_{1,2}$ & 93173.77 & & -4.44 & & 7 & \rdelim\}{4}{4mm} & \multirow{4}{*}{$1.183\pm 0.022$} & \multirow{4}{*}{$1.106\pm 0.052$}\\
					& $1_{2,1}\rightarrow 0_{1,0}$ & 93173.96 & & -4.95 & & 3\\
					& $1_{2,1}\rightarrow 0_{1,1}$ & 93173.96 & & -4.63 & & 3\\
					& $1_{2,1}\rightarrow 0_{1,2}$ & 93173.96 & & -5.84 & & 3\\
					& $1_{0,1}\rightarrow 0_{1,0}$ & 93176.26 & & -5.19 & & 3 & \rdelim\}{3}{4mm} & \multirow{3}{*}{$0.827\pm 0.022$} & \multirow{3}{*}{$0.461\pm 0.052$}\\
					& $1_{0,1}\rightarrow 0_{1,1}$ & 93176.26 & & -5.07 & & 3\\
					& $1_{0,1}\rightarrow 0_{1,2}$ & 93176.26 & & -4.67 & & 3\\
					\midrule
    			\bottomrule
			\end{tabular}
			}
			\label{tab:lines5}
\end{table}
\vspace*{\fill}

\newpage

\section{CS spectra and line fitting with {\scshape Radex}.}

Here we show the observed CS lines (see Table \ref{tab:summarylines}) and the synthetic spectra obtained with the radiative transfer code {\scshape Radex} and the parameters shown in Table \ref{tab:col_dens}.

\begin{figure*}[h]
	\centering
	\includegraphics[width=0.97\textwidth,keepaspectratio]{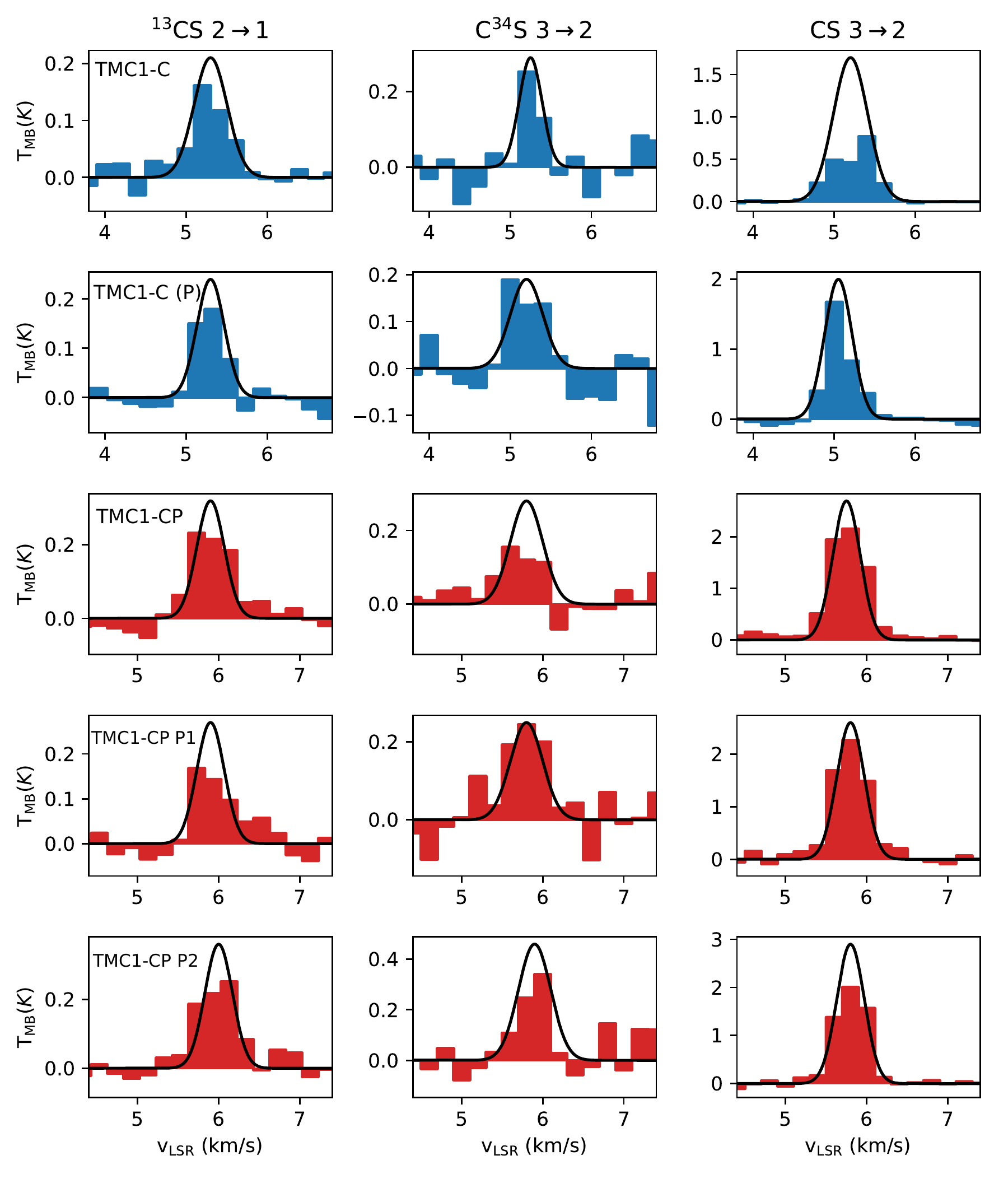}
	\caption{Observed $^{13}$CS $2\rightarrow 1$, C$^{34}$S $3\rightarrow 2$, and CS $3\rightarrow 2$ lines towards the positions listed in Table \ref{tab:targets} (TMC~1-C in blue and TMC~1-CP in red), with the synthetic spectra created with {\scshape Radex} and the parameters in Table \ref{tab:col_dens}.}
	\label{fig:csRadex}
\end{figure*}

\newpage

\section{HFS-CLASS results.}
The HFS-CLASS method is used to derive the opacities of the DCN, N$_{2}$H$^{+}$, and N$_{2}$D$^{+}$ lines from their hyperfine structure. We list here the results of this method applied at the observed lines of our setup.
\\
\begin{table}[!hb]
		\centering
		\caption{HFS-CLASS method results for the DCN, N$_{2}$H$^{+}$, and N$_{2}$D$^{+}$ lines at the observed positions. $T_{\rm A}$ is the line temperature of the main component, $\Delta V$ is the line width, and $\tau_{\rm m}$ is the optical depth of the main component.}
			\begin{tabular}{llcccc}
				\toprule
				\multirow{2}{*}{\textbf{Position}} & \multirow{2}{*}{\textbf{Species}} & \multirow{2}{*}{\textbf{Transition}} &  \multirow{2}{*}{\textbf{$T_{\rm A}\times\tau_{\rm m}$ (K)}} & \multirow{2}{*}{\textbf{$\Delta V$ (km s$^{-1}$)}} & \multirow{2}{*}{$\tau_{\rm m}$} \\
					
					& & & & & \\ \midrule \midrule 
					\multirow{4}{*}{TMC~1-C} & \multirow{2}{*}{DCN} & $1\rightarrow 0$ & $1.36\pm 0.17$ & $0.40\pm 0.02$ & $5.72\pm 0.84$\\
					& & $2\rightarrow 1$ & $0.25\pm 0.09$ & $0.37\pm 0.07$ & $3.24\pm 1.53$ \\ \cmidrule{2-6}
					& N$_{2}$H$^{+}$ & $1\rightarrow 0$ & $2.70\pm 0.08$ & $0.33\pm 0.02$ & $3.31\pm 0.17$ \\
					& N$_{2}$D$^{+}$ & $1\rightarrow 0$ & $0.33\pm 0.01$ & $0.50\pm 0.01$ & $0.10\pm 0.04$\\
					\midrule
					\multirow{4}{*}{TMC~1-C (P)} & \multirow{2}{*}{DCN} & $1\rightarrow 0$ & $1.57\pm 0.13$ & $0.34\pm 0.01$ & $3.45\pm 0.32$ \\
					& & $2\rightarrow 1$ & $0.35\pm 0.15$ & $0.43\pm 0.09$ & $3.19\pm 2.05$\\ \cmidrule{2-6}
					& N$_{2}$H$^{+}$ & $1\rightarrow 0$ & $2.67\pm 0.07$ & $0.33\pm 0.01$ & $2.64\pm 0.14$\\
					& N$_{2}$D$^{+}$ & $1\rightarrow 0$ & $0.25 \pm 0.01$ & $0.50\pm 0.01$ & $0.10\pm 0.02$\\
					\midrule
					\multirow{4}{*}{TMC~1-CP} & \multirow{2}{*}{DCN} & $1\rightarrow 0$ & $0.62\pm 0.08$ & $0.59\pm 0.03$ & $0.69\pm 0.35$\\
					& & $2\rightarrow 1$ & $0.18\pm 0.07$ & $0.38\pm 0.08$ & $0.41\pm 0.51$\\ \cmidrule{2-6}
					& N$_{2}$H$^{+}$ & $1\rightarrow 0$ & $2.49\pm 0.07$ & $0.44\pm 0.05$ & $1.65\pm 0.10$\\
					\midrule
					\multirow{4}{*}{TMC~1-CP (P1)} & \multirow{2}{*}{DCN} & $1\rightarrow 0$ & $0.85\pm 0.13$ & $0.49\pm 0.03$ & $1.14\pm 0.44$\\
					& & $2\rightarrow 1$ & $0.23\pm 0.04$ & $0.33\pm 0.12$ & $0.10\pm 0.73$\\ \cmidrule{2-6}
					& N$_{2}$H$^{+}$ & $1\rightarrow 0$ & $2.48\pm 0.09$ & $0.42\pm 0.06$ & $1.27\pm 0.11$\\
					\midrule
					\multirow{4}{*}{TMC~1-CP (P2)} & \multirow{2}{*}{DCN} & $1\rightarrow 0$ & $0.86\pm 0.14$ & $0.52\pm 0.04$ & $1.07\pm 0.49$\\
					& & $2\rightarrow 1$ & $0.50\pm 0.20$ & $0.50\pm 0.04$ & $2.22\pm 1.28$\\ \cmidrule{2-6}
					& N$_{2}$H$^{+}$ & $1\rightarrow 0$ & $2.38\pm 0.01$ & $0.42\pm 0.03$ & $1.34\pm 0.06$\\
					\midrule
    			\bottomrule
			\end{tabular}
			\label{tab:hfs}
\end{table}
\vspace*{\fill}

\newpage

\section{Collapse models, abundance predictions, and observations.}
In this section we show the observed abundances of DNC, DCN, CS, N$_{2}$H$^{+}$, and N$_{2}$D$^{+}$ at the positions in Table \ref{tab:targets}, and their predicted values according to the collapse models BES1, BES4, and AD presented in Section 9.

\vspace*{\fill}

\begin{figure*}[!hb]
	\centering
	\includegraphics[width=\textwidth,keepaspectratio]{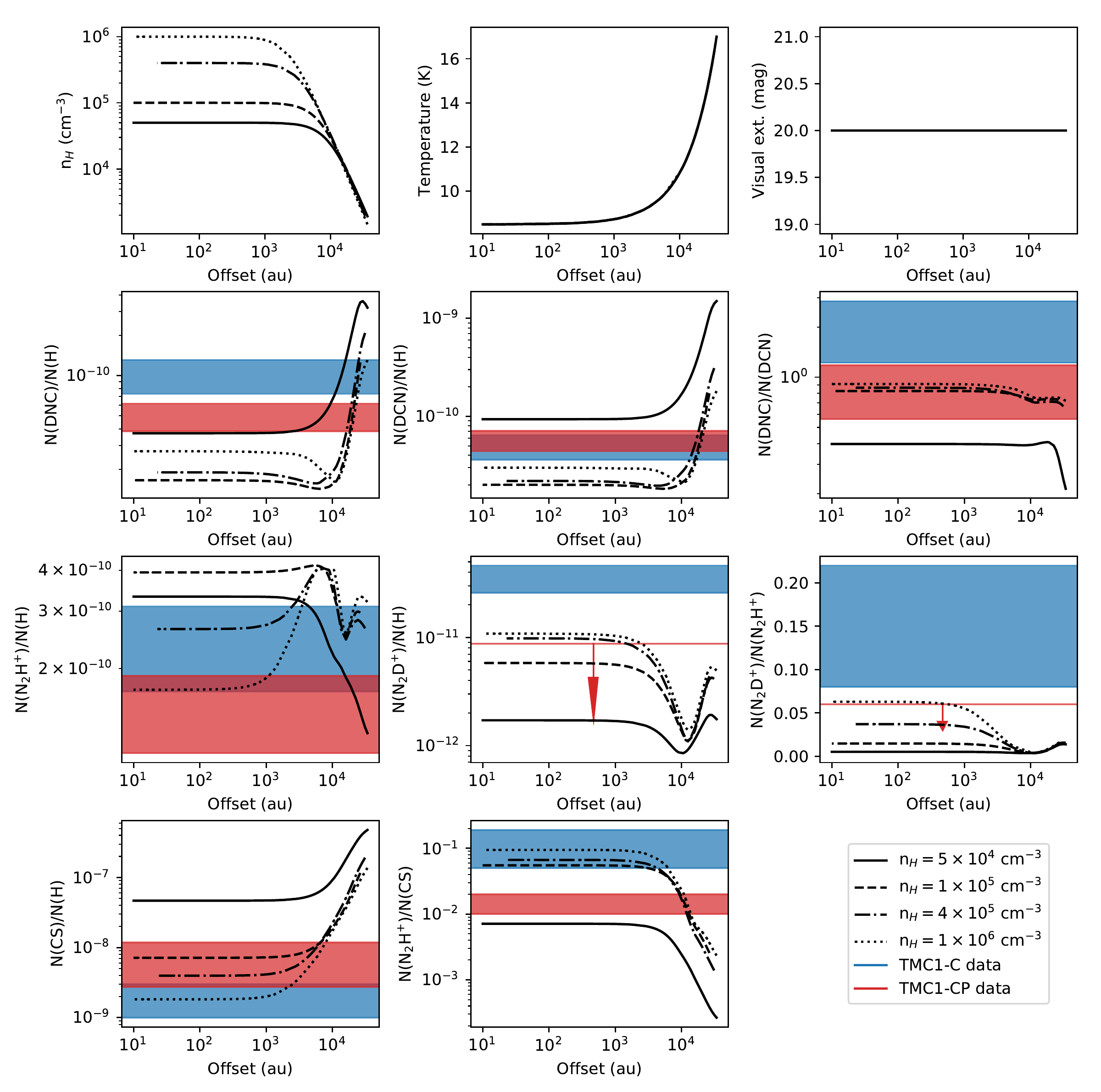}
	\caption{Predicted abundances along the line of sight (black) of the different BES1 collapse models with densities $5\times 10^{4}$ (solid), $10^{5}$ (dashed), $4\times 10^{5}$ (dot-dashed), and $1\times 10^{6}$ cm$^{-3}$ (dotted). The observational data at the center of the starless cores are plotted in blue (TMC~1-C) and red (TMC~1-CP). The arrows indicate upper bounds.}
	\label{fig:bes1}
\end{figure*}

\vspace*{\fill}
\newpage
\vspace*{\fill}

\begin{figure*}[!hb]
	\centering
	\includegraphics[width=\textwidth,keepaspectratio]{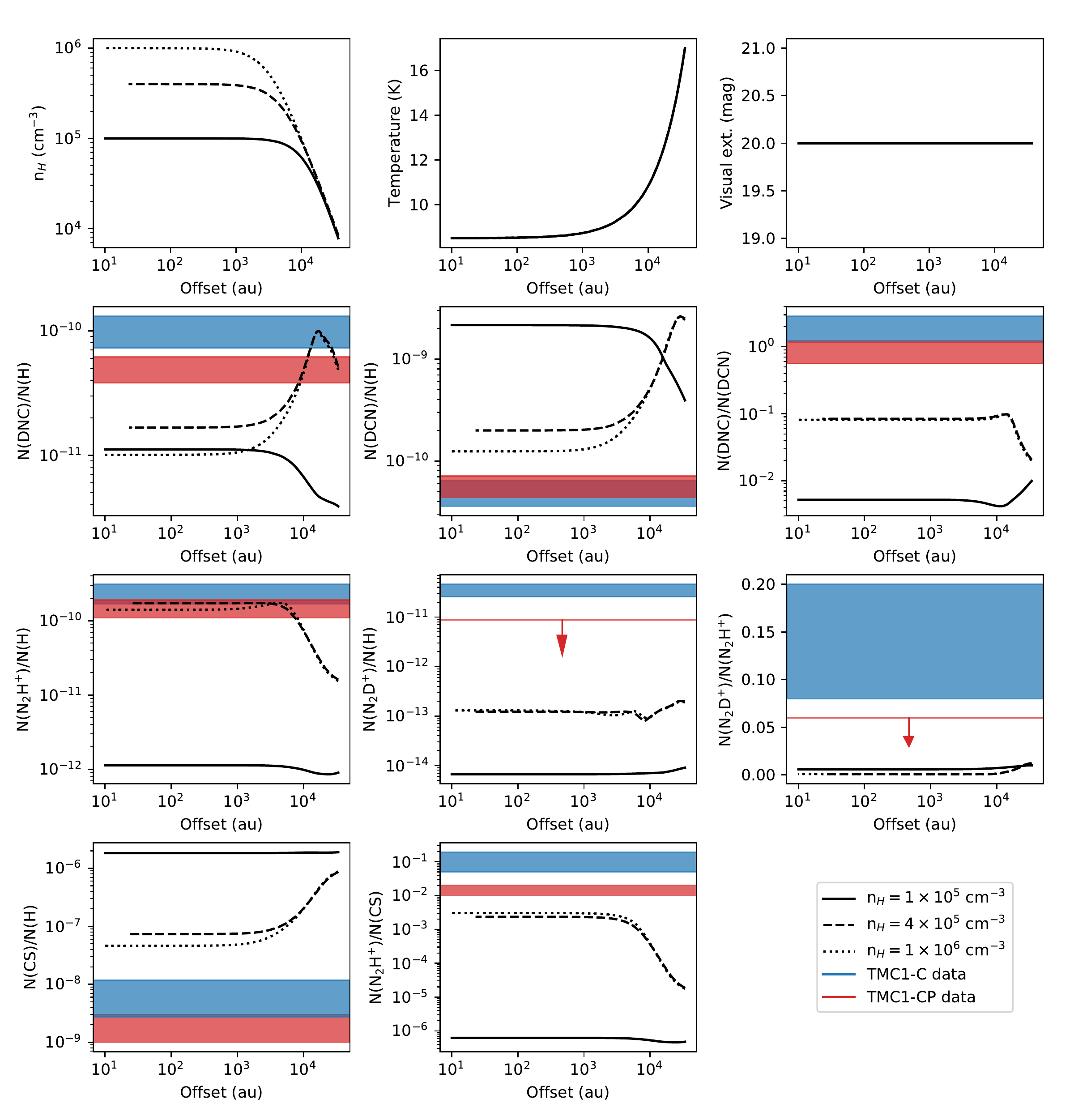}
	\caption{Predicted abundances (black) of the different BES4 collapse models with densities $10^{5}$ (solid), $4\times 10^{5}$ (dashed), and $1\times 10^{6}$ cm$^{-3}$ (dotted). The observational data at the center of the starless cores are plotted in blue (TMC~1-C) and red (TMC~1-CP). The arrows indicate upper bounds.}
	\label{fig:bes4}
\end{figure*}
\vspace*{\fill}
\newpage
\vspace*{\fill}
\begin{figure*}[!hb]
	\centering
	\includegraphics[width=\textwidth,keepaspectratio]{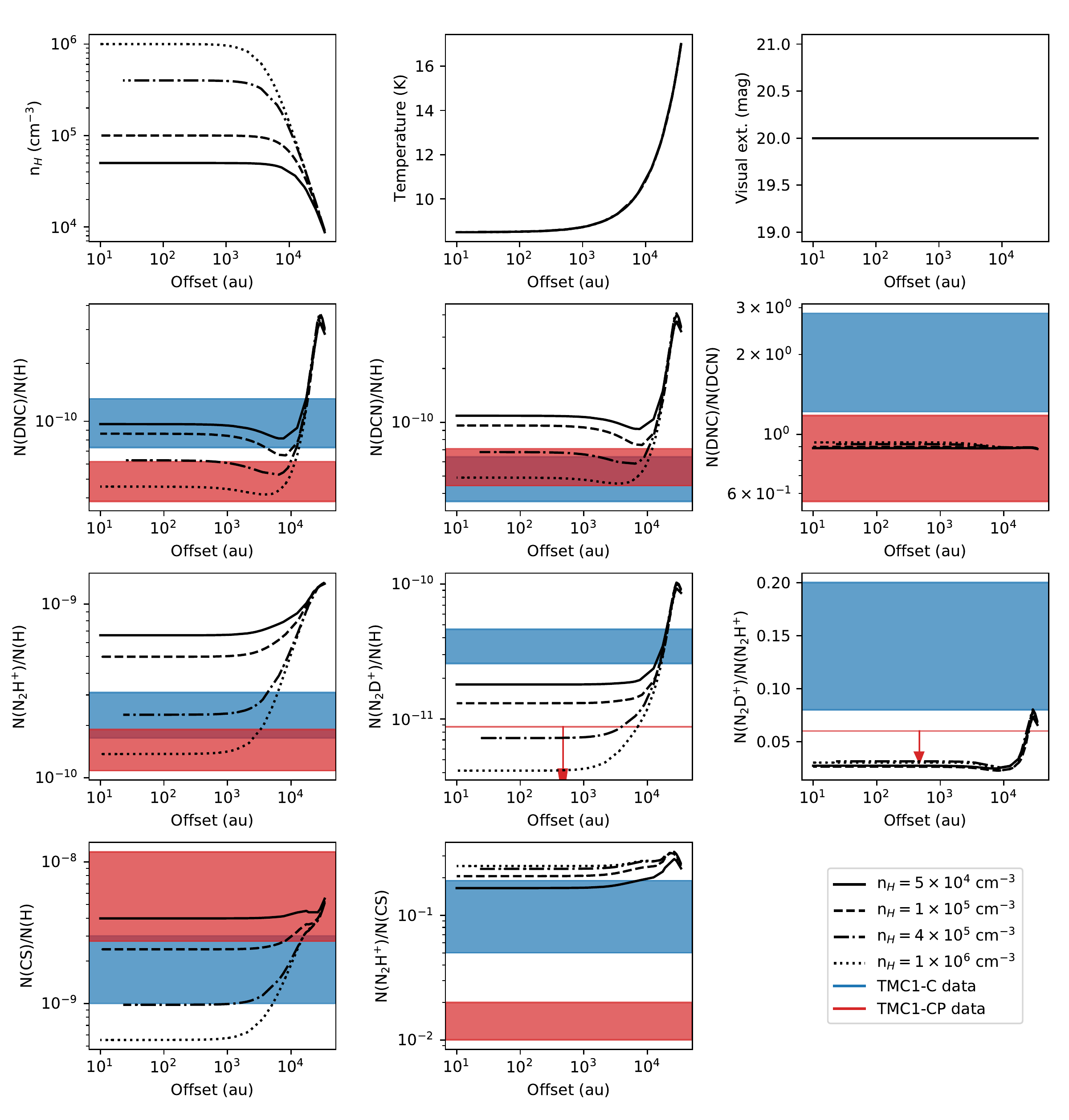}
	\caption{Predicted abundances (black) of the different AD collapse models with densities $5\times 10^{4}$ (solid), $10^{5}$ (dashed), $4\times 10^{5}$ (dot-dashed), and $1\times 10^{6}$ cm$^{-3}$ (dotted). The observational data at the center of the starless cores are plotted in blue (TMC~1-C) and red (TMC~1-CP). The arrows indicate upper bounds.}
	\label{fig:ad}
\end{figure*}
\vspace*{\fill}
\end{appendix}
\end{document}